\documentclass[twocolumn,twoside,pra,aps,showpacs,amsmath,amssymb,a4paper,superscriptaddress]{revtex4}

\usepackage{graphicx}
\usepackage{dcolumn}
\usepackage{bm}

\begin{document}

\preprint{APS}

\title{Quantum Scattering in Quasi-1D Cylindrical Confinement}

\author{J.I. Kim}
\affiliation{Physikalisches Institut, Universit\"{a}t Heidelberg, Philosophenweg 12, 69120
Heidelberg, Germany}
\author{J. Schmiedmayer}
\affiliation{Physikalisches Institut, Universit\"{a}t Heidelberg, Philosophenweg 12, 69120
Heidelberg, Germany}
\author{P. Schmelcher}
\affiliation{Physikalisches Institut, Universit\"{a}t Heidelberg, Philosophenweg 12, 69120
Heidelberg, Germany}
\affiliation{Theoretische Chemie, Institut f\"{u}r Physikalische Chemie, Universit\"{a}t
Heidelberg, Im Neuenheimer Feld 229, 69120 Heidelberg, Germany}


\date{June 21, 2005}

\begin{abstract} 
Finite size effects alter not only the energy levels of small systems, but can also lead 
to new effective interactions \emph{within} these systems. Here the problem of low energy 
quantum scattering by a spherically symmetric short range potential in the presence of a 
general cylindrical confinement is investigated. A Green's function formalism is developed 
which accounts for the full 3D nature of the scattering potential by incorporating \emph{all} 
phase-shifts and their couplings. This quasi-1D geometry gives rise to scattering
resonances and weakly localized states, whose binding energies and wavefunctions can be
systematically calculated. Possible applications include e.g. impurity scattering in
ballistic quasi-1D quantum wires in mesoscopic systems and in atomic matter wave
guides. In the particular case of parabolic confinement, the present formalism can also be
applied to pair collision processes such as two-body interactions. Weakly bound pairs and 
quasi-molecules induced by the confinement and having zero or higher orbital angular
momentum can be predicted, such as $p$- and $d$-wave pairings.  
\end{abstract}

\maketitle

\section{Introduction}
\label{intro}
Matter at very small dimensions can be strongly affected by boundary and surface effects. 
Bulk 3D properties may then change substantially or even disappear. Reaching this regime
poses a long term challenge motivated among others by the continuous technological drive
towards ever smaller information processing devices~\cite{itrs}. Below a certain limit of 
small length scales, novel quantum electronic properties may appear and be useful in order
to develop alternative devices~\cite{itrs-erd}. 

Recent developments in atomic physics have addressed similar questions by using 
atom-optical devices (or atom-chips) made out of surface-patterned 
substrates~\cite{weinstein1995a,folman2002a,reichel2002a,hinds1999a,fortagh2003a,wang2005a}
or by means of laser manipulation of atoms for trapping~\cite{grimm2000a} and atomic 
lithography~\cite{oberthaler2003a}. Analogous to electronic matter waves, these devices 
and techniques have the potential to trap and manipulate atomic matter waves down to 
sub-micron scales. A vast range of phenomena with atoms can then be studied, several of
them bearing close relationships to specific solid state systems. 

Of particular interest for both bulk and atomic systems is the quasi-1D regime of quantum
coherent \emph{single mode} transport. In this ultimate limit, the degrees of freedom in
the transversal trapping directions are effectively frozen in the quantum-limited ground
state of the confining potential. Such one-dimensional system can show distinct physical
properties e.g in quasi-1D quantum wires in two-dimensional electron gas (2DEG)
systems~\cite{itrs-erd,datta1997a,ferry1997a,chu1989a,bagwell1990a,gurvitz1993a,
bardarson2004a,indlekofer2005a} and  
Tomonaga-Luttinger liquids~\cite{tomonaga1950a,luttinger1963a,voit1994a}. In elongated
gases of ultracold atoms, the one dimensionality is revealed e.g. in magnetic  
guides~\cite{weinstein1995a,folman2002a,reichel2002a,hinds1999a,fortagh2003a,wang2005a},
in phase fluctuations of quasi-condensates~\cite{petrov2001a,goerlitz2001a,dettmer2001a}, 
the Tonks-Girardeau gas of impenetrable 
bosons~\cite{tonks1936,girardeau1960,lenard1966,tolra2004a,paredes2004a,kinoshita2004a} or  
one dimensional optical lattices~\cite{moritz2003a,moritz2005a}. In this broad context, we
address two processes that are affected by the quasi-1D geometry. On one hand, the matter
waves can be scattered off e.g. impurities or defects fixed along the guide. This
\emph{scattering by central fields} is an important variable in the transport properties
and can be strongly affected by the confinement. This has been revealed by several studies
of mesoscopic systems in the context of the Landauer approach to the quantized conductance
of quantum point-contacts in the presence of
impurities~\cite{chu1989a,bagwell1990a,gurvitz1993a,bardarson2004a}. On the other 
hand, two-body \emph{collision processes} should be dealt with when the matter wave
consists of interacting particles. A full understanding of these low dimensional systems
should therefore deal with both types of scattering. It should also derive \emph{ad hoc}
phenomenological parameters such as effective low dimensional coupling strengths from 
the real 3D character of the interactions. Another context in which such dimensional
reduction plays a role is the transmission and reflection of
electromagnetic~\cite{bostroem1981a} and elastic~\cite{olsson1994a} waves in transmission 
lines or resonators. 

New light has recently been shed on this subject by studies of atomic collisions in
1D~\cite{olshanii1998a,moore2004a,granger2004a,bergeman2003} and 
2D~\cite{petrov2000b,petrov2001b,petrov2004a}. Focusing on a parabolic cylindrical 
confinement, for which the center of mass can be eliminated, pioneering predictions for
collision processes are possible, e.g. that of confinement induced \emph{scattering
resonances} (CIR) in 1D. In such a resonance~\cite{olshanii1998a,moore2004a,granger2004a},
a total reflection may take place between the colliding atoms. The $s$-wave (or
zero-range) approximation to the scattering potential~\cite{olshanii1998a,moore2004a} may 
be partially lifted, provided the coupling of orbital angular momenta due to the
cylindrical geometry are not considered~\cite{granger2004a}. Another property induced by 
the confinement and closely related to CIR is the formation of weakly bound
\emph{quasi-molecules} with zero ($s$-wave) orbital angular momentum~\cite{bergeman2003},
which have been recently observed with singlet fermionic atoms in tight laser
traps~\cite{moritz2005a}. Coupling to the center of mass under non-parabolic confinement 
in the $s$-wave approximation is treated in~\cite{peano2005a} and can lead to the
appearance of additional CIRs. In the independent and distinct context of impurity
scattering studies, rectangular geometries predominate which are most suitable for the
growth of semiconductor heterostructures. Similar properties of resonant scattering and
weakly localized states are then predicted for non-interacting 2DEG systems. This is
demonstrated in the $s$-wave approximation to the scattering potential in~\cite{chu1989a}.
Finite range scatterers under general rectangular confinement are also considered, but in
an already reduced two-dimensional space~\cite{bagwell1990a,gurvitz1993a,bardarson2004a}. 

\begin{figure}
\includegraphics[scale=0.26]{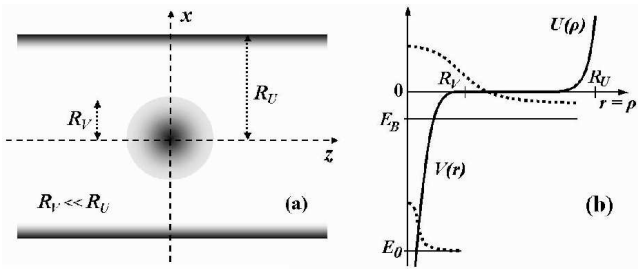} 
\caption{\label{guideeigenstates} ({\bf a}) Longitudinal profile of the cylindrical
guide. The scattering potential $V(r)$ is limited to within the spherical shaded area at
the center. ({\bf b}) Total potential energy along the transverse $xy$-plane passing
through the origin $\bm{r}=0$ (thick curve). For an attractive $V(r)$, one deep $E_0$ and
another shallow $E_B$ bound state without confinement are indicated, with their respective 
probability density profiles (dotted curves).} 
\end{figure}

In this work, a detailed formalism is presented that extends the results for impurity
scattering and collisions. It can treat general cylindrical confinement, parabolic or not,
and incorporate all scattering phase-shifts beyond $s$-waves, as well as the full
couplings of orbital angular momenta due to the broken spherical symmetry. A comprehensive
assessment of the scattering process reveals the most important mechanisms and parameters
at play. From this unified description, CIRs are seen to be a general low energy effect in
quasi-1D geometries. It might be useful e.g. as an alternative gating mechanism in low
power transistor-like devices that could incorporate quasi-1D 
structures~\cite{itrs-erd,datta1997a,ferry1997a,chu1989a,bagwell1990a,gurvitz1993a,bardarson2004a, 
indlekofer2005a,effmass}. The existence of weakly localized bound states and 
quasi-molecules can, in turn, be extended to include higher orbital angular momentum
pairing, whose binding energies \emph{and} wavefunctions can be systematically computed.  

This paper is organized as follows. In section~\ref{lowd} the present problem of 
scattering in confined geometries is formulated for both scattering by central fields and
collision processes. In section~\ref{cirpseudo}, the $s$-wave approximation to the
scattering potential is briefly reviewed. In section~\ref{phaseshifts}, after an overview
of the free space scattering, our formalism in terms of Green's functions and phase-shifts 
is developed. In section~\ref{resonantscatt}, this formalism is applied to analyze two
types of confinement induced effects, namely, CIR and weakly localized states. A
discussion of the main approximations used is given in section~\ref{harmonic}, followed by 
the conclusions.

\section{Scattering in Low Dimensionality}
\label{lowd} 
In a low energy scattering process under confinement, one expects the scattering potential
to affect primarily the degrees of freedom not experiencing external confinement. The other
degrees of freedom are then forced to occupy only one or very few confined states. However, 
the very existence of these confined states do play a role due to virtual transitions. The
effects of the latter can become noticeable when properly examining the unconfined degrees
of freedom. In this regard one is lead to calculate effective scattering amplitudes as
given by Eq.(\ref{f1d2}). 

These amplitudes provide a complete description of the quasi-1D problem. For real
energies, they describe the scattering properties of running waves. Weakly localized
states are then found in the complex-energy plane.

\subsection{Hamiltonian under confinement}
\label{cf}
Consider first the scattering of a \emph{single} particle of mass $\mu$ and coordinates
$\bm{r}=(x,y,z)$ by a central field \emph{fixed} at the origin. Let $V(r)$ be the
spherically symmetric scattering potential and $U(\rho)$ the cylindrically symmetric
confining potential, where $\bm{\rho}=(x,y)$ denotes the transverse coordinates. The
Hamiltonian is then given by 
\begin{equation}
\label{hrelative}
H = -\frac{\hbar^2}{2\mu}\nabla^2 + U(\rho) + V(r)\, .
\end{equation}
If $R_V$ and $R_U$ are the ranges of $V(r)$ and $U(\rho)$, respectively, then $|V|\ll V_0$
for $r\gg R_V$ and $U\ll U_0$ for $\rho\ll R_U\,$, where $V_0$ and $U_0$ are
characteristic energy scales, such as the respective ground state energies (see
fig.~\ref{guideeigenstates}). Two particular examples are the parabolic confinement 
$U(\rho)=\mu\,\omega_\perp^2\rho^2/2$, for which $R_U$ is of the order of the oscillator
length $a_\perp=(\hbar/\mu\,\omega_\perp)^{1/2}$ and the square-well (or hard-wall) such
that $U(\rho)\rightarrow+\infty$ for $\rho\geq R_U$ and zero otherwise. The first condition
on the scattering and on the confining potentials assumed in this work is $R_V\ll
R_U\,$. In this sense, both the potentials $V(r)$ and $U(\rho)$ can be relatively
\emph{general} for this type of scattering process. This condition implies that there is a
distinct region $R_V\ll r\ll R_U$ where spherical symmetry still prevails and one can
define scattering phase-shifts. This is an (intermediate) asymptotic region, where the
effect of the confinement $U(\rho)$ is felt only as a boundary condition to the scattering
by $V(r)$ (see section~\ref{decompose}). Fig.~\ref{guideeigenstates} depicts the geometry
of the problem. Another important condition for the validity of the present approach to 
scattering in confined geometries is the condition $k\sim 1/R_U$ of low total scattering
energy $E\equiv \hbar^2k^2/2\mu$ (or Fermi energy) for the Hamiltonian $H$. Only few
excited transverse states $\varphi_n$ above the ground state (see Eq.(\ref{eigentrans}))
can be effectively populated. The scattering process is thus assumed to occur at low
temperatures or under sufficiently tight confinement.  

Alternatively, consider the case of \emph{collisions} between e.g. two cold atoms of
masses $m$ and coordinates $\bm{r}_i=(\bm{\rho}_i,z_i)$,
$i=1,2\,$. $\bm{R}=(\bm{r}_1+\bm{r}_2)/2$ and $\bm{r}=\bm{r}_1-\bm{r}_2$ denote the center 
of mass (CM) and relative coordinates, respectively. The total Hamiltonian is  
\begin{equation*}
\label{h}
H_2 = \sum_{i=1}^2\left[ -\frac{\hbar^2}{2m}\nabla^2_i + U_c(\rho_i) \right] + V(r)\, ,
\end{equation*}
where $U_c$ is the cylindrical confinement and $V$ the two-body interaction potential. For
this scattering process $U_c$ should be \emph{parabolic}, so that the CM can be separated  
exactly. The Hamiltonian for the relative coordinates is then given by
Eq.(\ref{hrelative}), where $\mu=m/2$ becomes the (collisional) effective mass and 
$U(\rho)=2\,U_c(\rho/2)\,$. For non-parabolic confinement, additional scattering 
resonances that can occur even at weak coupling to the CM~\cite{peano2005a} is not
included in Eq.(\ref{hrelative}). However, at ultracold temperatures for which only the 
lowest transversal states of the guide are occupied, the probability to find the atoms in
the center of the guide is largest. Consequently, it might be a reasonable zeroth-order
approximation to assume on average that the CM coordinates are also confined to the center 
of the guide. Therefore Eq.(\ref{hrelative}) is a starting point to the dynamics of the relative 
coordinates for a general interaction $V(r)$, in which the CM fluctuations around the mean
value $\langle \bm{R}\rangle=0$ are at the present stage neglected.  
 
Because of the symmetry of $U(\rho) + V(r)$, the scattering solutions can be restricted to 
be axially symmetric. The corresponding axially symmetric and orthonormalized eigenstates
of the confined variables $\varphi_n(\rho)$, $n=0,1,2,\dots$, satisfy 
\begin{equation}
\label{eigentrans}
\left[-\frac{\hbar^2}{2\mu}\left(\frac{\partial^2}{\partial x^2} 
			+ \frac{\partial^2}{\partial y^2}\right) + U(\rho)\right] 
				\varphi_n(\rho)=\epsilon_n\, \varphi_n(\rho) 			
\end{equation}
where $\epsilon_n\equiv \hbar^2q_n^2/2\mu>0$ are the transversal energies, with $q_0\sim 
1/R_U\,$.

\subsection{1D scattering}
\label{1dscatt}
In a purely 1D problem let the 1D interaction potential $V_{1D}(z)$ have the finite range
$R_{1D}$ around the origin $z=0$, such that $V_{1D}\approx 0$ for $|z|\gg R_{1D}$. The 
scattering problem for an incoming plane wave $e^{ik_0z}$ can be described by the
amplitudes $f^{\pm}$ in the asymptotic behaviour of the solution $\psi(z)$ in the
transmission and reflection regions 
\begin{equation*}
\label{asymp1d}
\psi(z) = \left\{ \begin{array}{ll}
			e^{ik_0z} + f^+e^{ik_0z}\, , & R_{1D}\ll z, \\
			e^{ik_0z} + f^-e^{-ik_0z}\, ,    & z\ll -R_{1D}.
	  	  \end{array}
	  \right.
\end{equation*}
Current conservation then yields 
\begin{equation}
\label{conserv1d}
|1+f^+|^2 = 1 - |f^-|^2\, .
\end{equation}
Particular cases are resonant transmission ($f^+\rightarrow 0$) and total reflection
($f^+\rightarrow -1$) at finite $k_0$ and finite potentials.

\subsection{Effective 1D scattering}
\label{eff1dscatt}
The above 1D picture corresponds to a fixed unperturbed transversal state, e.g., the
ground state $\varphi_0(\rho)$. The total wave function would then have the form
$\psi(z)\varphi_0(\rho)$ where scattering effects would occur only in $\psi(z)$. However,
transitions to other states $\varphi_n(\rho)$, $n=1,2,\dots$, even as virtual transitions,
arise from the coupling between the $\varphi_n$'s caused by the scattering potential
$V(r)$. 

Let $\Psi(\bm{r})$ be the full 3D axially symmetric scattering solution to
Eq.(\ref{hrelative}), 
\begin{subequations}
\begin{equation}
\label{hrelativ2}
\left[\nabla^2 - u(\rho) + k^2\right]\Psi(\bm{r}) = v(r)\,\Psi(\bm{r}),
\end{equation}
where $E\equiv \hbar^2k^2/2\mu>0$ is the total scattering energy, $u(\rho)\equiv 2\mu 
U(\rho)/\hbar^2$ and $v(r)\equiv 2\mu V(r)/\hbar^2$. The above mentioned couplings and 
virtual transitions are best seen by expanding the solution in the cylindrical basis
defined by the $\varphi_n$'s, 
\begin{equation}
\label{expand}
\Psi(\bm{r})=\sum_{n=0}^\infty\,\psi_n(z)\,\varphi_n(\rho),
\end{equation}
and solving for each $\psi_n(z)$. Substituting into Eq.(\ref{hrelativ2}), multiplying by
$\varphi_n^\ast$, integrating and using Eq.(\ref{eigentrans}) yields 
\begin{equation*}
\left( \frac{d^2}{dz^2} + k^2 - q_n^2 \right)\psi_n(z) = 
	\int dx\,dy\, \varphi_n^\ast(\rho) v(r) \Psi(\bm{r})\, .
\end{equation*}
For a given total energy $E$, one defines the longitudinal wave vectors
$k_n$ by 
\begin{equation*}
k_n = \left\{ \begin{array}{ll}
	        \sqrt{k^2 - q_n^2}\,, & \hspace{2em} \mbox{if $0\leq n\leq n_E$}, \\ 
	        i\sqrt{q_n^2 - k^2}\,, & \hspace{2em} \mbox{if $n > n_E$},
	      \end{array}
	\right.
\end{equation*}
where $n_E$ is the largest integer such that $q_{n_E}<k<q_{1+n_E}$ and denotes the highest
``open channel''. In the single mode regime $n_E=0$. As usual, the general solution
$\psi_n(z)$ can be expressed in terms of a homogeneous solution and a particular
inhomogeneous solution, $\psi_{f,n}(z)$. This inhomogeneous solution can be written in
terms of the 1D Green's function 
\begin{equation*}
G_n(z)=-\frac{e^{ik_n|z|}}{2ik_n}\, , \hspace{1em} G_n''+ k_n^2G_n = - \delta(z-z'),
\end{equation*}
describing an outward scattering for $n\leq n_E$ and an exponentially decaying virtual
state for the ``closed channels'' $n> n_E$. Then for all $n$ 
\begin{equation*}
\psi_{f,n}(z) = - \int d^3\bm{r}'\, 
	G_n(z-z')\,\varphi_n^\ast(\rho')\, v(r')\,\Psi(\bm{r}')\, .
\end{equation*}
The expansion Eq.(\ref{expand}) of $\Psi(\bm{r})$ takes then the form  
\begin{equation}
\label{psin}
\Psi(\bm{r}) 
	     = \sum_{n=0}^{n_E}\,b_n\,e^{ik_nz}\varphi_n(\rho) 
		 + \sum_{n=0}^\infty\,\psi_{f,n}(z)\,\varphi_n(\rho)\, ,
\end{equation}
\end{subequations}
where the homogeneous part $b_n\,e^{ik_nz}$ of each $\psi_n(z)$, for some constant
$b_n$, gives rise to the total incoming state and thus is limited to the open channels
$n\leq n_E$ only.  

The asymptotic condition is obtained in the limit $|z|\rightarrow\infty$ by neglecting the
exponentially decaying terms $\psi_{f,n}(z)$ in Eq.(\ref{psin}) for $n>n_E$. Using then
$|z-z'|=|z|\mp z'$ in $G_n$, for $z\rightarrow\pm\infty$, one has 
\begin{subequations}
\label{f1dall}
\begin{equation}
\label{f1d1}
\Psi(\bm{r}) \approx
\sum_{n=0}^{n_E}\left[b_ne^{ik_nz} + f_n^\pm\,e^{ik_n|z|}\right]\varphi_n(\rho)\, ,
\end{equation}
where the {\em effective 1D scattering} amplitudes $f_n^\pm$ are defined by (for $n\leq 
n_E$) 
\begin{equation}
\label{f1d2}
f_n^\pm \equiv  \frac{1}{2ik_n}\int d^3\bm{r'}\left[e^{\pm ik_nz'}
	\varphi_n(\rho')\right]^\ast v(r')\,\Psi(\bm{r}'),
\end{equation}
for forward $z>0$ and backward $z<0$ scattering, respectively.
\begin{figure}
\includegraphics[scale=0.27]{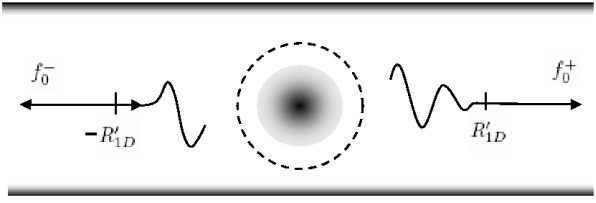} 
\caption{\label{effR1D} A pictorial view of a scattering ``path'' of a particle coming
from the left. Outside the effective 1D range $|z|\gg R'_{1D}$, the particle is
stabilized by the confinement back into e.g. the tansversal ground state (for the single
mode regime $n_E=0$). Here the net effect of the scattering is contained in the amplitudes
$f_0^-$ and $f_0^+$ of backward and forward scattering, respectively. Within the dashed
circle, however, the spherical symmetry prevails. The phase information of the wave
function within this region is then captured by $f_0^\pm$.} 
\end{figure}
Note first that although $|z|\gg |z'|$, the phase $k_nz'$ in $f_n^\pm$ is not necessarily
negligible since $k_n|z'|\sim 2\pi$ for sufficiently large momentum $k_n$ and depending on
the range $R_V$ of the interaction $v(r')$. This phase is negligible only when the
wavelength $2\pi/k_n$ of the incoming wave is not able to resolve the details of the potential
$v(r')$ at low momenta. In this case, a zero-range or $s$-wave approximation to $v(r')$ is 
valid. However, this approximation should break down when the scattering occurs at short 
wavelengths or when higher angular momenta become necessary. 
Note that there is no clearly defined 1D range $R_{1D}$ as in the purely 1D case, since
the asymptotics given by Eq.(\ref{f1d1}) depends on the 
convergence behaviour of the series in Eq.(\ref{psin}). If the convergence is slow, one
expects an {\em effective 1D range}, say $R'_{1D}$, substantially larger than
$R_V$ or even $R_U$. It is an effective range for the 1D collision and is of the order of
the minimum separation from the scattering center, such that Eq.(\ref{f1d1}) is the
dominant contribution to the scattering wave function (see fig.~\ref{effR1D}). As regards
collision processes, this effective range sets some limits on the average equilibrium
distance between the colliding particles. 

The present formalism to quantum scattering under confinement is approximate but
non-perturbative, in the sense that there is no small parameter around which an expansion
is performed. The determination of its range of validity is chosen here by imposing the
conservation of probability on the final wavefunctions. 
The probability conservation condition in 3D and the identity 
$\int_\Omega d^3\bm{r}\,\mbox{div$\bm{j(\bm{r})}$}
=\int_{S_c+S_1+S_2}d^2\bm{S}\cdot\bm{j(\bm{r})}$
for a volume $\Omega$ enclosed by a large cylinder of surface area $S_c$ and by two
transversal discs of surface areas $S_1$ at $z_1<0$ and $S_2$ at $z_2>0$ yields 
\begin{eqnarray*}
0 &=& \int_{S_c} d\phi\,dz\,\rho\,\left[\hat{\bm{\rho}}\cdot\bm{j}(\bm{r})\right] \\
  & & \hspace{2em}	- \int_{S_1} dx\,dy\left[\hat{\bm{k}}\cdot\bm{j}(\bm{r})\right] 
			+ \int_{S_2} dx\,dy\left[\hat{\bm{k}}\cdot\bm{j}(\bm{r})\right], 
\end{eqnarray*} 
where $\hat{\bm{\rho}}$ and $\hat{\bm{k}}$ are cylindrical unit vectors. Applying 
$\Psi=\sum_n\psi_n\varphi_n$, the first integral on the rhs vanishes for large radius
$\rho\rightarrow\infty$ since each $\varphi_n(\rho)\rightarrow 0$. The other two integrals
yield 
\[ 
0 = \sum_{n=0}^\infty\left[ -j_n(z_1) + j_n(z_2)\right], 
\]
where $j_n(z)$ are 1D currents calculated with $\psi_n(z)$. Setting $z_{1,2}\sim \mp
R'_{1D}$ one can use the asymptotics for $\psi_n$ in Eq.(\ref{f1d1}) so that the total
conservation condition reads 
\begin{equation}
\label{conservation}
\hspace{-0.15em}
0=\sum_{n=0}^{n_E}\left(|b_n+f^+_n|^2 + |f^-_n|^2 - |b_n|^2\right)k_n\, .
\end{equation}
\end{subequations}
This equation extends the pure 1D result in Eq.(\ref{conserv1d}). It also determines how
the initial flux is distributed among the open channels. In other words, even when the
initial state occupies only a single channel, i.e., $b_n=\delta_{0,n}$, new channels are
occupied due to scattering, if the total energy is large enough $n_E>0$. This conservation
condition can serve then as a gauge for the range of validity of our approximations to the 
amplitudes $f^{\pm}_n$.

\section{Scattering in the $s$-wave approximation}
\label{cirpseudo}
In many circumstances, the low energy scattering in free space can be well described by
approximating the interaction $V(r)$ by Fermi-Huang's zero-range
potential~\cite{huang1987,fetter1971a}  
\begin{subequations} 
\begin{equation}
\label{pseudo}
V_\delta(r) = \frac{2\pi\hbar^2a}{\mu}\,\delta(\bm{r})
				\frac{\partial}{\partial r}(r\ \cdot\ )\, .
\end{equation} 
This approximation singles out the $l=0$ angular momentum component of the collision and
thus depends only on the $s$-wave scattering length $a$. The regularization $\partial(r\ 
\cdot\ )/\partial r$ assures that the singularity $1/r$ of the full scattering solution to
this $V_\delta(r)$ is properly dealt with. 

For collisions in a strong parabolic confinement and using Eq.(\ref{pseudo}), the series
in Eq.(\ref{psin}) also has a singularity for short distances $r\rightarrow 0$ very
similar to $1/r$. By properly dealing with this singularity, with~\cite{olshanii1998a} or
without any regularization~\cite{moore2004a}, the amplitudes $f_n^\pm$ in Eq.(\ref{f1d2})
can be calculated. In the single mode regime ($n_E=0$) and for small longitudinal momenta 
$k_0$ one obtains 
\begin{equation}
\label{f0pseudo}
f_0^\pm = - \frac{1}{
	1+i\left[-\,\frac{a_\perp^2}{2a}\left(1-\frac{Ca}{a_\perp}\right)\right]k_0 
		   }\,,	
\end{equation}
\end{subequations}
where $C=1.4603\dots$ and $a_\perp$ is the harmonic oscillator length of the parabolic
confinement~\cite{olshanii1998a,moore2004a}. Here, a total reflection $f_0^+ = -1$ is
predicted for large scattering lengths, on the order of the confinement length, such that 
$a_\perp=Ca\,$, which is called the confinement induced resonance
(CIR)~\cite{olshanii1998a}. This total reflection means that the colliding pair
experiences a diverging effective 1D interaction along the longitudinal cylindrical axis
(see section~\ref{bosons}). 

It will be seen later that such a singularitiy should actually
be the \emph{same} singularity $1/r$. It arises from the 3D free space Green's function
and becomes dominant as $r\rightarrow 0\,$ away from the confining boundaries (see
sections~\ref{gfunctions} and~\ref{harmonic}). This latter identification is one of the
elements that will allow us to incorporate all partial waves $l\geq 1$ besides the
$s$-wave. In addition, the above conclusions concerning CIR under parabolic confinement
will be seen to hold not only for general interactions $V(r)$, as first examined 
in~\cite{granger2004a}, but also for general \emph{non-parabolic} confinement
$U(\rho)\,$. In this sense the present formalism is of a more general character.

\section{Scattering Phase-Shifts}
\label{phaseshifts}
In the following, a general formalism is developed in order to calculate the scattering
amplitudes for any given finite short-ranged and spherically symmetric scattering 
potential $V(r)$. This is done by expressing the amplitudes $f_n^\pm$ in terms of the
scattering phase-shifts $\delta_l$ associated to this $V(r)$. Our formalism is also able
to deal with a broad range of confinements $U(\rho)$ as well as scattering energies above
the transversal ground state. Particularly relevant is the straightforward accounting of
angular momenta couplings due to the confinement. We begin by providing the connection to
scattering in free space.

\subsection{Phase shifts in free space scattering}
\label{freescatt}
The standard formalism for free space scattering by a central potential is recast in a
form suitable for later comparison. The main idea is to express all quantities relevant 
for scattering in terms of the phase-shifts. The rationale is that these phase-shifts can
be considered as intrinsic to a given scattering potential $V(r)$, without much regard to 
details of external boundary conditions.  

In the absence of any confinement, the scattering solution can be written as 
\begin{subequations}
\begin{equation}
\label{3dwf}
\Phi(\bm{r}) = \Phi_i(\bm{r}) 
	- \int d^3\bm{r}'\, G(\bm{r},\bm{r}')\, v(r')\, \Phi(\bm{r'}),
\end{equation}
where $\Phi_i(\bm{r})$ is the incoming state and $G(\bm{r},\bm{r}')$ is the free space
Green's function. In terms of outward and inward scattered waves, the latter reads
\begin{equation}
\label{3dgf}
G(\bm{r},\bm{r}') = \gamma_+\,\frac{e^{ik|\bm{r}-\bm{r}'|}}
				{4\pi|\bm{r}-\bm{r}'|}  
		    + \gamma_-\,\frac{e^{-ik|\bm{r}-\bm{r}'|}}
				{4\pi|\bm{r}-\bm{r}'|}, 
\end{equation}
for some constants $\gamma_{\pm}$ obeying $\gamma_+ + \gamma_- = 1$. A single constant
$\gamma$ can be introduced by setting
\begin{equation}
\label{3dgamma}
\gamma_{\pm}\equiv \frac{1\pm\gamma}{2}.
\end{equation}
\end{subequations}
In free space scattering, the inward component proportional to $\gamma_-$ is usually
absent (i.e., $\gamma=1$). However, one expects this \emph{inflow} of particles if there
is e.g. an exterior confinement that forces the scattered particles back towards the
center. This interpretation will be validated in section~\ref{gfunctions} (see also
Fig~\ref{effR1D}). $\Phi_i(\bm{r})$ and $G(\bm{r},\bm{r}')$ are determined by the boundary
conditions imposed on $\Phi(\bm{r})$, e.g. that of an axially symmetric solution. 

The scattering variables of interest here
are the amplitudes $f_{\mathrm{out},\mathrm{in}}$ of outward and inward waves, scattered
along a direction given by $\theta$ and $\phi$ at large distances $r\gg R_V$ from the
scattering center. Using the expansion $|\bm{r}-\bm{r}'|\sim r-\bm{r}\cdot\bm{r}'/r$ in
Eq.(\ref{3dgf}), Eq.(\ref{3dwf}) becomes  
\begin{subequations}
\label{3dasympamp}
\begin{equation}
\label{3dasymp}
\Phi(\bm{r}) \approx \Phi_i(\bm{r}) + f_{\mathrm{out}}\frac{e^{ikr}}{r} 
		+ f_{\mathrm{in}}\frac{e^{-ikr}}{r}\, ,  \hspace{1em} r\gg R_V\, ,
\end{equation}
with
\begin{equation}
\label{3damp}
f_{\mathrm{out},\mathrm{in}} = -\frac{\gamma_{\pm}}{4\pi}\int d^3\bm{r}'\, 
		\left[e^{\pm i\bm{k}\cdot\bm{r}'}\right]^\ast v(r')\,\Phi(\bm{r}'),
\end{equation}
\end{subequations}
where $\bm{k}\equiv k\bm{r}/r$ depends on the direction $(\theta,\phi)$ of $\bm{r}$. It 
can be clearly seen that these amplitudes have a shape similar to their equivalent
effective 1D amplitudes $f_n^{\pm}$ in Eq.(\ref{f1d2}). To express the amplitudes in terms
of the phase-shifts an angular momentum decomposition of $f_{\mathrm{out},\mathrm{in}}$ is 
necessary. Following standard procedures~\cite{morse1953,mott1965}, we obtain from
Eq.(\ref{3damp}) 
\begin{subequations}
\label{3damp2tau}
\begin{equation}
\label{3damp2}
f_{\mathrm{out},\mathrm{in}} = -\gamma_{\pm}\sum_{l=0}^\infty 
	(\mp i)^l(2l+1)\,\tau_l\,P_l(\cos{\theta})\, ,
\end{equation}
with the definition
\begin{equation}
\label{3dtau}
\tau_l\equiv \frac{1}{4\pi}\int d^3\bm{r}'
		\left[j_l(kr')\,P_l(\cos{\theta'})\right]\,v(r')\,\Phi(\bm{r}')\, ,
									\hspace{1em}
\end{equation}
\end{subequations}
where $j_l$ and $P_l$ are the spherical Bessel functions and Legendre polynomials,
respectively. The next step is to relate the  elements $\tau_l$ to the
phase-shifts. Decompose $\Phi_i(\bm{r})$ and $G(\bm{r},\bm{r}')$ in spherical
coordinates. Substituting then into Eq.(\ref{3dwf}) for $r>r'\sim R_V$ and carrying out
the $\phi'$-integration one obtains for $r> R_V$
\begin{eqnarray*}
\label{3dsph2}
\Phi(\bm{r}) &=& 
	\sum_{l=0}^\infty (2l+1)(\beta_l - ik\gamma\tau_l)\,j_l(kr)\,P_l(\cos{\theta})
								\nonumber \\
	& & \hspace{1em}
	    + \sum_{l=0}^\infty (2l+1)(\beta'_l + k\tau_l)\,n_l(kr)\,P_l(\cos{\theta})\, ,
\end{eqnarray*}
having used Eq.(\ref{3dgamma}) and Eq.(\ref{3dtau}) and the constants $\beta$ and $\beta'$
are related to $\Phi_i(\bm{r})$. On the other hand, the phase-shifts arise when directly
decomposing this same solution $\Phi(\bm{r})$ in spherical coordinates and looking at the
radial solutions outside the interaction region $r>R_V$, 
\begin{equation*}
\label{3dsph}
\Phi(\bm{r}) = \sum_{l=0}^\infty c_l\left[\cos{\delta_l}\,j_l(kr) 
		- \sin{\delta_l}\,n_l(kr)\right] P_l(\cos{\theta}). 
\end{equation*}
Here the phase-shifts $\delta_l$ are calculated from the behaviour of the radial parts
$R_l(r)$ inside the scattering region $r<R_V$ only, irrespective of the details of the
outer boundary conditions~\cite{mott1965}. The boundary conditions are matched by properly 
choosing the constants $c_l$, i.e. by superposing the spherical components. Comparing
these last two forms of $\Phi(\bm{r})$ 
yields, 
\begin{subequations}
\label{3drelate}
\begin{eqnarray}
\label{3drelate1}
c_l &=& (2l+1)\frac{\beta_l + i\gamma\,\beta'_l}
				{\cos{\delta_l} - i\gamma\,\sin{\delta_l}}\, , \\
\label{3drelate2}
\tau_l  &=&  \frac{1}{k}\, \frac{\beta_l\sin{\delta_l} + \beta'_l\cos{\delta_l}}
				{\cos{\delta_l} - i\gamma\,\sin{\delta_l}}\, ,
					\hspace{1.5em} l=0,1,\dots\hspace{0.5em}
\end{eqnarray}
\end{subequations}
Considering the phase-shifts $\delta_l$ (or the scattering potential $v(r)$) as given, 
the second relation Eq.(\ref{3drelate2}) solves for the amplitudes 
$f_{\mathrm{out},\mathrm{in}}$ in Eq.(\ref{3damp2}) (and also the integral equation
Eq.(\ref{3dwf}) at distances $r>R_V$). The first relation Eq.(\ref{3drelate1}) in turn
determines how to superpose the spherical components in  $\Phi=\sum_l c_lR_lP_l$
in order to match the boundary condition contained in Eq.(\ref{3dasymp}).

\subsection{Green's functions under confinement}
\label{gfunctions} 
Just as the free space 3D scattering amplitudes in Eqs.(\ref{3dasympamp}) can be written
in terms of the phase-shifts via Eqs.(\ref{3damp2tau}) and Eq.(\ref{3drelate2}), a very
similar procedure can be applied to calculate the amplitudes $f_n^\pm$ characterizing the
1D effective dynamics. In this section a key element of the formalism is worked out,
namely, the Green's function $G_c$ under confinement. It is shown that $G_c$ can be
expressed in terms of the free space Green's function $G$ as in Eq.(\ref{gcgu2}). The
final purpose, to be achieved in section~\ref{decompose}, is to express it as a series 
in spherical coordinates, valid at least in the center of the guide where the spherical
symmetry prevails over the cylindrical one due to the main assumption of a short range
scattering potential $v(r)$ and a confining potential $u(\rho)$ that is sufficiently 
flat at the center, $R_V\ll R_U$. Such a series allows then the straightforward
introduction of the phase-shifts. 

Consider the solution $\Psi$ to Eq.(\ref{hrelativ2}) given by the expansion in
Eq.(\ref{psin}). One can rewrite Eq.(\ref{psin}) in the integral form (keeping again the 
$\phi'$-integration as in Eq.(\ref{3dtau}) for convenience) 
\begin{subequations}
\begin{equation}
\label{int}
\Psi(\bm{r}) = \Psi_i(\bm{r})-\int d^3\bm{r}' 
		G_c(\bm{r},\bm{r}') v(r')\Psi(\bm{r}')\, ,
\end{equation}
where the incident state is 
\begin{equation}
\label{intin}
\Psi_i(\bm{r}) = \sum_{n=0}^{n_E}b_n\,e^{ik_nz}\varphi_n(\rho)\, ,
\end{equation}
and $G_c$, the axially symmetric Green's function of the confining potential, has the form 
\begin{equation}
\label{gc}
G_c(\bm{r},\bm{r}') =
\sum_{n=0}^\infty\varphi_n(\rho)\varphi^\ast_n(\rho')\,G_n(z-z')\, .
\end{equation}
\end{subequations}
It is convenient to introduce a non-axially symmetric Green's function $G_u$ under the
confining potential $u(\rho)$. This $G_u$ then satisfies 
$
\left[\nabla^2 - u(\rho) + k^2\right]G_u(\bm{r},\bm{r}')= -\delta(\bm{r}-\bm{r}')
$.
If the $\phi$-independent solution $\Psi(\bm{r})$ is written as an integral equation 
using this $\phi$-dependent $G_u$, it follows from comparison with the integral equation
Eq.(\ref{int}) that $G_c$ and $G_u$ are related by 
$
2\pi\, G_c(\bm{r},\bm{r}') = \int d\phi'\, G_u(\bm{r},\bm{r}')
$.
Compare now the differential equations of $G_u$ and that of the free space Green's
function $G$ of Eq.(\ref{3dgf}) in the region of the scattering potential such that
$r,r'\ll R_U$. Since $u(\rho)\approx 0$ by assumption, they should differ at most by a 
homogeneous term, say $\Delta_u$, so that  
$
G_u(\bm{r},\bm{r}') \approx G(\bm{r},\bm{r}') +
		\Delta_u(\bm{r},\bm{r}')$,
for $r,r'\ll R_U$,
where $\Delta_u$ satisfies the homogeneous equation $(\nabla^2 +
k^2)\Delta_u(\bm{r},\bm{r}')=0$ and we obtain 
\begin{equation}
\label{gcgu2}
\hspace{-0.1em}
G_c(\bm{r},\bm{r}') \approx \int\frac{d\phi'}{2\pi} G(\bm{r},\bm{r}')
		+ \Delta_c(\bm{r},\bm{r}'),    \hspace{0.8em}    r,r'\ll R_U,
							\hspace{-0.3em}
\end{equation}
where $\Delta_c\equiv \int d\phi'\Delta_u/2\pi$. In order to obtain the function
$\Delta_c$ and $\gamma$ in Eq.(\ref{3dgamma}) in terms of the scattering parameters, we
first write the closed channel ($n>n_E$) wavefunctions of $G_c$ in Eq.(\ref{gc}) and
$G$ in Eq.(\ref{3dgf}) using cylindrical Bessel functions. The latter represents a good
approximation to these transversal eigenstates for a given potential $U(\rho)$ close to
the guide center $\rho\ll R_U$ where $U(\rho)\approx 0$. By subsequently comparing $G_c$
and $G$ we can determine $\Delta_c$. 

First we express the transversal eigenstates $\varphi_n(\rho)$ in terms of the normalized
Bessel function according to 
\begin{subequations}
\begin{equation}
\label{besselapprox}
\varphi_n(\rho) \approx \frac{N_n}{\pi^{1/2}R_U}J_0(q_n\rho)\, , 
					\hspace{1.5em} \rho\ll R_U,\  n>n_E,
\end{equation}
where $N_n=|J_1(r_{n+1})|^{-1}$ and $r_{n+1}$ is the $(n+1)$-th root of $J_0$, related to
the transverse momenta $q_n$ by $q_nR_U=r_{n+1}$. A good expression for the roots
$r_{n+1}$ is given by the cosine approximation $J_0(x)\approx (2/\pi
x)^{1/2}\cos{(x-\pi/4)}$. This leads to 
\begin{equation}
\label{besselroot}
q_n\approx \left(n + \frac{3}{4}\right)\frac{\pi}{R_U}, \hspace{2em} n>n_E\,.
\end{equation}
\end{subequations}
The corresponding energy dispersion relation $\epsilon_n\sim n^2$ is quadratic with
respect to $n$ compared to the linear one $\epsilon_n\sim 2n$ for, e.g., 2D harmonic
oscillators (see e.g.~\cite{olshanii1998a}). This should not cause any confusion since
this excited spectrum $n>n_E$ is summed over in Eq.(\ref{gc}) and leads to the confinement
independent free space Green's function $G$ in Eq.(\ref{gcgu2}). It is the lower part of
the spectrum $n\leq n_E$, in turn, that relates to the term $\Delta_c$.   

Indeed, we substitute Eq.(\ref{besselapprox}) in the series for the closed channels in
Eq.(\ref{gc}) yielding for $r,r'\ll R_U$ 
\begin{eqnarray*}
\lefteqn{\hspace{-1.5em}G_c(\bm{r},\bm{r}')
   \approx \sum_{n=0}^{n_E} \varphi_n(\rho)\,\varphi^\ast_n(\rho')\, 
			   \frac{e^{i\sqrt{k^2-q_n^2}|z-z'|}}{2(-i)\sqrt{k^2-q_n^2}}
	}							\nonumber\\
 \hspace{2em} 
 &+&\sum_{n=1+n_E}^\infty \frac{q_n\,\delta q_n}{4\pi}\,J_0(q_n\rho)J_0(q_n\rho')
				\,\frac{e^{-\sqrt{q_n^2-k^2}|z-z'|}}{\sqrt{q_n^2-k^2}},
								\nonumber\\
\end{eqnarray*}
where $\delta q_n=\pi/R_U$ is the increment of $q_n$ and we used $|J_1(r_{n+1})|^2=2/\pi
q_nR_U$. For $r,r'\ll R_U$ a continuum approximation to the closed channel series can now
be applied. Separating the real and imaginary parts of $G_c$ (supposing $\varphi_n(\rho)$
real for $n\leq n_E$ without much loss of generality) gives 
\begin{eqnarray}
\label{gcgu4}
\lefteqn{\hspace{-1.5em}G_c(\bm{r},\bm{r}')}			\nonumber\\
   &\hspace{-1.5em}\approx& 
	\hspace{-0.5em}i\sum_{n=0}^{n_E} \varphi_n(\rho)\,\varphi_n(\rho')\,
  		\frac{\cos{\left(\sqrt{k^2-q_n^2}|z-z'|\right)}}{2\sqrt{k^2-q_n^2}}	
								\nonumber\\
   & & \hspace{0.5em}  - \sum_{n=0}^{n_E} \varphi_n(\rho)\,\varphi_n(\rho')\,
 	\frac{\sin{\left(\sqrt{k^2-q_n^2}|z-z'|\right)}}{2\sqrt{k^2-q_n^2}}	
								\nonumber\\
   & & \hspace{0.5em}  + \int_{q_{1+n_{E}}}^\infty \frac{q\,dq}{4\pi}J_0(q\rho)J_0(q\rho')
			\frac{e^{-\sqrt{q^2-k^2}|z-z'|}}{\sqrt{q^2-k^2}},
\end{eqnarray}
where $q_{1+n_E}>k$ designates the transversal momentum of the first ``virtual'' or closed
channel $n=1+n_E$. As for $G$, one uses now the identity~\cite{morse1953}  
\begin{eqnarray*}
\frac{e^{ik|\bm{r}-\bm{r}'|}}{4\pi|\bm{r}-\bm{r}'|} & = & 
	  - \sum_{m=0}^\infty 
		\left(2-\delta_{0,m}\right)\cos{[m(\phi-\phi')]} \nonumber\\
	& &	      \times\int_0^\infty \frac{q\,dq}{4\pi}J_m(q\rho)J_m(q\rho')
				\frac{e^{i\sqrt{k^2-q^2}|z-z'|}}{i\sqrt{k^2-q^2}},
								  \nonumber\\
\end{eqnarray*}
and the correct branch $0\leq\mathrm{Arg}\sqrt{k^2-q^2}<\pi$ in order to obtain for the
integrated free space Green's function   
\begin{eqnarray}
\label{gcyl}
\lefteqn{\int\frac{d\phi'}{2\pi}\, G(\bm{r},\bm{r}')				
	}								\nonumber\\
   & = & i\gamma\int_0^k \frac{q\,dq}{4\pi}J_0(q\rho)J_0(q\rho')\,		
	    \frac{\cos{\left(\sqrt{k^2-q^2}|z-z'|\right)}}{\sqrt{k^2-q^2}} \nonumber\\
   &   & \hspace{0.5em} - \int_0^k \frac{q\,dq}{4\pi}J_0(q\rho)J_0(q\rho')\,		
	    \frac{\sin{\left(\sqrt{k^2-q^2}|z-z'|\right)}}{\sqrt{k^2-q^2}} \nonumber\\
   &   & \hspace{1em} + \int_k^\infty \frac{q\,dq}{4\pi}J_0(q\rho)J_0(q\rho')
		\frac{e^{-\sqrt{q^2-k^2}|z-z'|}}{\sqrt{q^2-k^2}}.  
\end{eqnarray}
To compare with the Green's function $G_c$ in Eq.(\ref{gcgu4}), we specialize
Eq.(\ref{gcyl}) to $r,r'\ll R_U$ and impose the condition for low total scattering
energy. 

The value of $\gamma$ follows from the comparison of the imaginary parts of
Eqs.(\ref{gcgu4}) and (\ref{gcyl}). Changing the integration variable from $q$ to
$\sqrt{k^2-q^2}$ in the first integration on the rhs of Eq.(\ref{gcyl}) and using the low
energy condition in the imaginary parts of Eq.(\ref{gcgu4}) and Eq.(\ref{gcyl}), $\gamma$
can be identified in the limit $r,r'\rightarrow 0$  
\begin{subequations}
\label{gammakdeltac}
\begin{equation*}
\gamma = 2\pi\sum_{n=0}^{n_E}\frac{|\varphi_n(0)|^2}{k\,\sqrt{k^2-q_n^2}}\, .
\end{equation*}
In other words, since $r,r'\ll R_U\sim 1/k\,$, both imaginary parts are approximated by a 
single and the same constant $\gamma k/4\pi$, independent of \mbox{\boldmath{$r$} and
\boldmath{$r'$}}. Although exact values of $|\varphi_n(0)|^2$ can be employed for a given
potential $U(\rho)$, it is more instructive to use general approximations such as
$|\varphi_n(0)|^2\sim 1/\pi R_U^2$ obtained through normalization. An estimate that takes
into account the spatial variation and the few nodes of $\varphi_n$ is the square well
approximation, namely, $|\varphi_n(0)|^2\approx N_n^2/\pi R_U^2$. Therefore, without 
loss of generality, one can thus write 
\begin{equation}
\label{gammak}
\gamma = \sum_{n=0}^{n_E}\frac{2N_n^2}{R_U^2\,k\,\sqrt{k^2-q_n^2}}\, .
\end{equation}
It should be mentioned that one can improve on how these low energy poles $q_n$ of $G_c$
are treated. However, it suffices to use for now the above result in order to present the 
formalism. 

By comparing the real parts of Eq.(\ref{gcgu4}) and Eq.(\ref{gcyl}), it is seen that
$\Delta_c$ in Eq.(\ref{gcgu2}) is given by 
\begin{eqnarray*}
\lefteqn{\Delta_c(\bm{r},\bm{r}')} 						\\
	&=&  - \int_k^{q_{1+n_{E}}} \frac{q\,dq}{4\pi}J_0(q\rho)J_0(q\rho')
		\frac{e^{-\sqrt{q^2-k^2}|z-z'|}}{\sqrt{q^2-k^2}} 		\\
 &	& \hspace{2em} + \left[\int_0^k \frac{d\bar{p}}{4\pi}
		J_0(\bar{q}\rho)J_0(\bar{q}\rho')\,\sin{\left(\bar{p}|z-z'|\right)} 
	    \right.								\\
 &	& \hspace{3em} - \left.\sum_{n=0}^{n_E} \varphi_n(\rho)\,\varphi_n(\rho')\,
			 	\frac{\sin{\left(k_n|z-z'|\right)}}{2k_n}
	    		\right], 					
\end{eqnarray*}
where $k_n=\sqrt{k^2-q_n^2}\,$ and $\bar{q}\equiv\sqrt{k^2-\bar{p}^2}$. The first term on
the rhs stems from the offset between the lower limits of integration $q_{1+n_E}$ and $k$
in Eq.(\ref{gcgu4}) and Eq.(\ref{gcyl}), respectively. It accounts for the discreteness of
the low lying transversal states due to the confinement. The second term in the square
brackets is of order $(k^2/8\pi - \sum_{n=0}^{n_E}N_n^2/2\pi R_U^2)\,|z-z'|$ and thus is
relatively smaller by a factor of $|z-z'|/R_U$ compared to the first term, whose order of
magnitude $(q_{1+n_E}^2-k^2)^{1/2}/4\pi\sim 1/R_U$ can be estimated by calculating the
integral for $\bm{r},\bm{r}'\sim 0\,$. Neglecting this second term yields 
\begin{equation}
\label{deltac}
\Delta_c(\bm{r},\bm{r}') = - \int_0^{p_c}
		\frac{dp}{4\pi}\,J_0(q\rho)J_0(q\rho')\,e^{-p|z-z'|}\, ,
\end{equation}
\end{subequations}
where $q\equiv(k^2+p^2)^{1/2}$ and $p_c\equiv(q_{1+n_E}^2-k^2)^{1/2}$. The relation
Eq.(\ref{gcgu2}) is then proved and will be the basis of an angular momenta 
decomposition of the effective amplitudes $f_n^\pm$ in section~\ref{decompose}. In this
regard, the weak dependence of $\Delta_c$ on the coordinates $\bm{r}$ and $\bm{r}'$ is
kept, since this dependence introduces couplings between orbital angular momenta. This
will be seen as $\Delta_c$ is decomposed in spherical coordinates in the following.

\subsection{Angular momenta decomposition}
\label{decompose}
We are now in the position to express the effective 1D scattering amplitudes
$f_n^\pm$ in terms of the scattering phase-shifts $\delta_l$ of a given scattering
potential. The first step is to decompose these amplitudes in Eq.(\ref{f1d2}) in a similar   
way as was done in the free space case in Eqs.(\ref{3damp2tau}) by introducing ``matrix''
elements similar to $\tau_l$. Subsequently these quantities are related to the
phase-shifts $\delta_l$ in a manner analogous to Eqs.(\ref{3drelate}). 

Applying the condition $R_V\ll R_U$, the transversal states $\varphi_n$ in the integrand
of the amplitudes Eq.(\ref{f1d2}) should be well approximated by the Bessel functions
Eq.(\ref{besselapprox}). This is because $r'\sim R_V$ whereas $|k_n|\sim q_n\sim
1/R_U\,$. It is enough then to find expansions of $e^{\pm ik_n z}J_0(q_n\rho)$ in
spherical coordinates, with $k^2=q_n^2+k_n^2$. For this purpose one may invert the
following identity~\cite{morse1953}  
\begin{eqnarray*}
P_l(\cos{\bar{\theta}})\, j_l(kr) &=& \frac{1}{2i^l}\int_0^\pi 
	dw\,\sin{w}\, e^{ikr\cos{\bar{\theta}}\cos{w}} \\
&&	   \hspace{3.5em} \times\, J_0(kr\sin{\bar{\theta}}\sin{w}) P_l(\cos{w})
\end{eqnarray*}
by using the orthonormality property of the Legendre polynomials. As a result, one obtains
for $r\ll R_U$ 
\begin{equation}
\label{alphaln} 
e^{ik_nz}\,\varphi_n(\rho) = \sum_{l=0}^\infty 
			i^l(2l+1)\,\alpha_{ln}\,j_l(kr)\,P_l(\cos{\theta})\, , 
\end{equation}
where $\alpha_{ln}\approx N_nP_l(k_n/k)/\pi^{1/2}R_U$. For more quantitative results, these
coefficients $\alpha_{ln}$ can be evaluated numerically from the exact eigenfunctions
$\varphi_n(\rho)$. One can then express the amplitudes $f_n^\pm$ in Eq.(\ref{f1d2}) as the
following series in angular momenta for $n\leq n_E$
\begin{subequations} 
\begin{eqnarray}
\label{f1d-spherical}
f_n^\pm 
	&=& f_{ng}\  \pm\  f_{nu} 				\\
	&\equiv& \sum_{l\,\mathrm{even}}\frac{4\pi\,(2l+1)\,\alpha_{ln}}{2ik_n}\,\,T_l\ 
		\pm \sum_{l\,\mathrm{odd}}\frac{4\pi\,(2l+1)\,\alpha_{ln}}{2ik_n}\,\,T_l
									\nonumber
\end{eqnarray}
having used the parity property $P_l(-x)=(-)^lP_l(x)$. The ``matrix'' element $T_l$ is
defined by  
\begin{equation}
\label{Tl}
T_l\equiv \frac{1}{i^l\,4\pi}\int d^3\bm{r}'\,
			[j_l(kr')\,P_l(\cos{\theta'})]\, v(r')\,\Psi(\bm{r}').
\end{equation}
\end{subequations}
The non-zero momenta $l\geq 1$ contributions to $f_n^\pm$ stem from the (small)
dependence on $z'$ and $\rho'$ of the integrand in the definition of $f_n^\pm$. Their
neglect amounts to assuming a point like zero range interaction $v(r)$, for which only the 
$s$-wave remains. In the free space scattering case, the corresponding quantity equivalent
to this $T_l$ is $\tau_l$ given in Eq.(\ref{3dtau}). The key difference now is that each
$T_l$ will depend on {\em all other} $T_l$'s, meaning that the confined scattering
solution $\Psi(\bm{r})$ in the definition of $T_l$ couples the angular momenta, while the
free space solution $\Phi(\bm{r})$ in the definition of $\tau_l$ does not. 
 
The next step is to relate the $T_l$ to the phase-shifts $\delta_l$. From the 
results in Eqs.(\ref{gammakdeltac}) for the approximation Eq.(\ref{gcgu2}), the Green's  
function $G_c$ can be decomposed into spherical coordinates in the region $r',r\ll R_U$ by 
separately decomposing $G$ and $\Delta_c$. The free space part $G$ can be decomposed 
following the procedures of section~\ref{freescatt}. Then, for $r'<r$, 
\begin{eqnarray}
\label{gexpanded}
\hspace{-1.5em}
\int\frac{d\phi'}{2\pi}\, G(\bm{r},\bm{r}') 
	&=& ik\sum_{l=0}^\infty j_l(kr') \left[ \gamma\,j_l(kr) + i\,n_l(kr) \right] 
										\nonumber\\ 
	& & \hspace{2em}
		\times\frac{2l+1}{4\pi}P_l(\cos{\theta})P_l(\cos{\theta'}),
\end{eqnarray}
having used the expansion of $e^{\pm ik|\bm{r}-\bm{r}'|}/4\pi|\bm{r}-\bm{r}'|$ in
spherical harmonics and $\gamma_\pm=(1\pm\gamma)/2$. To evaluate $\Delta_c$ we use the
analytic continuation $\bar{\theta}=\pi/2-i\theta'$ of the first identity of this section 
to obtain 
\begin{eqnarray*}
e^{-pz}J_0(q\rho) 
&=& \sum_{l=0}^\infty \left[i^l(2l+1)P_l(ip/k)\right] j_l(kr)P_l(\cos\theta)\, , 
\end{eqnarray*}
where $p=k\sinh{\theta'}$ and $q=k\cosh{\theta'}$ such that $k^2=q^2-p^2$. The expansion
of $\Delta_c$ in Eq.(\ref{deltac}) for $r,r'\ll R_U$ can then be written as 
\begin{eqnarray}
\label{dexpanded}
\Delta_c(\bm{r},\bm{r}') &=& 
	- \sum_{l=0}^\infty \left[ \int_0^{p_c}\frac{dp}{4\pi}
				P_l(\sigma_{zz'} ip/k)e^{\sigma_{zz'} pz'}J_0(q\rho')
			    \right] 				\nonumber\\
&&\hspace{4em}		\times\, i^l(2l+1) j_l(kr)P_l(\cos\theta)\, , 
\end{eqnarray}
where $\sigma_{zz'}\equiv \mathrm{sign}(z-z')$. Inserting Eq.(\ref{dexpanded}) and
Eq.(\ref{gexpanded}) back into Eq.(\ref{gcgu2}), the scattering solution $\Psi(\bm{r})$ in
Eq.(\ref{int}) becomes for $R_V\ll r \ll R_U$ 
\begin{eqnarray}
\label{spherical}
\Psi(\bm{r})\hspace{-0.2em} &\approx&\hspace{-0.2em} \sum_{l=0}^\infty
  i^l(2l+1) \left[ \alpha_l + \gamma_l(z) - i\gamma kT_l \right]j_l(kr)P_l(\cos{\theta}) 
								\nonumber\\
	& &	\hspace{1em} + \sum_{l=0}^\infty i^l(2l+1) 
			\left[\, kT_l \,\right] n_l(kr)P_l(\cos{\theta})\, ,
\end{eqnarray}
where $\alpha_l\equiv\sum_{n=0}^{n_E}b_n\alpha_{ln}$ comes from the expansion of 
$e^{ik_nz}\,\varphi_n(\rho)$ in the initial state $\Psi_i(\bm{r})$ in Eq.(\ref{intin}) and
$\gamma_l(z)$ is defined by
\begin{eqnarray*}
\label{}
\gamma_l(z) &\equiv& \int_0^{p_c}\frac{dp}{4\pi} \int d^3\bm{r}'	
		P_l(\sigma_{zz'}ip/k) 					\nonumber\\
 & & \hspace{5em} \times\,\left[e^{\sigma_{zz'} pz'}J_0(q\rho')\right]
							v(r')\Psi(\bm{r}'). 
\end{eqnarray*}
The $z$-dependence of this $\gamma_l(z)$ arises from the sign of $\sigma_{zz'}$ in
$\Delta_c$. Because of this $z$-dependence, the above decomposition of $\Psi(\bm{r})$ is
not fully spherically symmetric. It reflects the fact that an approximate radial solution
on the left side of the guide at some $z_1=-|z_0|$ needs not coincide necessarily with
one on the right side at $z_2=|z_0|$ since the problem has no perfect spherical symmetry. 

A better understanding of the role of this $\gamma_l(z)$ arises when one tries to write it
in terms of the $T_l$'s that determine the amplitudes $f_n^\pm$ in
Eq.(\ref{f1d-spherical}) and the solution in Eq.(\ref{spherical}). Expanding 
$e^{\sigma_{zz'} pz'}J_0(q\rho')$ (see above) in the integrand in the definition of
$\gamma_l(z)$ one obtains
\begin{eqnarray*}
\hspace{0em}\gamma_l(z)&=& 					
\sum_{s=0}^\infty (2s+1)\left[\int_0^{p_c}dp\,P_l(ip/k)P_s(ip/k)\right]\\
&& \hspace{-2em} \times\int d^3\bm{r}'\,\frac{(\sigma_{zz'})^{l+s}}{i^s\,4\pi}
	[j_s(kr')\,P_s(\cos{\theta'})]\, v(r')\,\Psi(\bm{r}'),		\nonumber
\end{eqnarray*}
assuming the summation and integration can be interchanged (see section~\ref{f0pm}). If it
happened that $(\sigma_{zz'})^{l+s}= 1$, then $\gamma_l(z)$ would be a  linear sum of the 
elements $T_s$ and the scattering solution in Eq.(\ref{spherical}) would have spherical 
symmetry. Such a constant $\gamma_l(z)$ could be possible if one could neglect in the 
series for $\gamma_l(z)$ terms for which the parity of $P_s$ is different from that of
$P_l$, such that for each $l$, only terms satisfying  
$l+s=\mathrm{even}$
would be retained. This condition can be related to the fact that the confining potential
$u(\rho)$ only couples angular momenta with the same parity, in the sense that 
$\langle l|\hat{u}|s\rangle\equiv 
\int d\theta'P_l(\cos{\theta'})u(r\sin{\theta'})P_s(\cos{\theta'})=0$ if
$l+s=\mathrm{odd}$. Therefore, a consistent approximation to the scattering solution
$\Psi(\bm{r})$ is found if one neglects odd parity combinations in the series defining
$\gamma_l(z)$ and sets $\gamma_l(z)\approx \gamma_l$, such that 
\begin{eqnarray}
\label{gammal}
\gamma_l(z) &\approx & \gamma_l						\nonumber\\
&\equiv& \sum_{s[l]} (2s+1)\left[k\int_0^{p_c/k}dx\, P_l(ix)P_s(ix)\right] T_s\nonumber\\
&\equiv& \sum_{s[l]} (2s+1)\,P_{ls}\,T_s\, , \hspace{1.5em} l=0,1,2,\dots\, ,
\end{eqnarray}
where $s[l]$ denotes the sum over even (odd) $s$ for even (odd) $l$.  

Using Eq.(\ref{gammal}), the scattering phase-shifts $\delta_l$ can now be introduced,
since the expansion in Eq.(\ref{spherical}) becomes spherically symmetric. Indeed, this
expansion can be conveniently rewritten by the introduction of constants $c_l'$ and
$\delta_l$ {\em defined}, for the moment, by ($l=0,1,2,\dots$)
\begin{subequations}
\label{definecl}
\begin{eqnarray}
c_l'\cos{\delta_l}&\equiv& 
		i^l(2l+1) \left[ \alpha_l + \gamma_l - i\gamma kT_l \right], \\
c_l'\sin{\delta_l}&\equiv&
		- i^l(2l+1)\,kT_l\,, 
\end{eqnarray}
\end{subequations}
such that our solution has the form 
\begin{equation}
\label{spherical-delta} 
\hspace{-0.25em}
\Psi(\bm{r}) \approx \sum_{l=0}^\infty
	c_l'\left[\cos{\delta_l}\,j_l(kr) - \sin{\delta_l}\,n_l(kr)\right]
								P_l(\cos{\theta}).
\end{equation}
In order to identify the $\delta_l$, one notes, on one hand, that this wavefunction
$\Psi(\bm{r})$ can be directly obtained as a spherically symmetric series 
$\sum_lc_lR_l(r)P_l(\cos{\theta})$ in the region $r\ll R_U$, where $u(\rho)$ is
negligible, if one solves the Schr\"odinger equation moving outward from the origin
$\bm{r}=0$ (see Fig~\ref{effR1D}). The radial part $R_l$ can then be separately determined 
and for $R_V\ll r\ll R_U$ it should be given by Eq.(\ref{spherical-delta}), provided one
chooses $c_l=c_l'$. On the other hand, the solution for the scattering potential $v(r)$
without confinement is given by a series with the {\em same} radial parts $R_l$ but {\em
different} constants $c_l\neq c_l'$ (see Section~\ref{freescatt}). The asymptotics of
$R_l$ in this free space case is then also given by Eq.(\ref{spherical-delta}) so that
$\delta_l$ are indeed the phase-shifts of the unconfined scattering problem. The
difference between the confined and free space solutions, in the region $R_V\ll r\ll R_U$,
is thus accounted for by the different constants $c_l'$ and $c_l$, respectively, related
to distinct boundary conditions. Besides, $\delta_l$ depends solely on the solution of
$R_l$ in the interior region $r<R_V$~\cite{mott1965}. The above defining relations for
$c_l'$ and $\delta_l$ in Eqs.(\ref{definecl}) are then \emph{equations} that determine
$c_l'$ and $T_l$ in terms of $\delta_l$, namely,  
\begin{equation*} 
c_l'= \frac{(2l+1)(\alpha_l+\gamma_l)\,i^l}
		{\cos{\delta_l} - i\gamma \sin{\delta_l}},  \hspace{2em}
T_l = \frac{\alpha_l+\gamma_l}{i\gamma k - k\cot{\delta_l}}\, ,
\end{equation*}
Finally, using Eq.(\ref{gammal}) one obtains the matrix equation relating $T_l$ to the
full ensemble of 3D free space scattering phase-shifts $\delta_l$ ($l=0,1,2,\dots$)
\begin{equation}
\label{tmatrix}
\left(i\gamma k - k\cot{\delta_l}\right)T_l = \alpha_l + \sum_{s[l]} (2s+1)P_{ls}T_s\, .
\end{equation}
The coupling of angular momenta brought about by the confinement is a result of
$\Delta_c$, which accounts for the confining geometry and the discreteness of the low
lying transversal states that should be resolved at low energies. The main equations that
allow us the analysis of the effective 1D scattering in confined geometries are 
Eqs.(\ref{f1dall}), Eq.(\ref{f1d-spherical}) and Eq.(\ref{tmatrix}). The probability
conservation condition Eq.(\ref{conservation}) serves to gauge the range of validity of
the results. Although the above result is enough to present the formalism, improvements 
to Eq.(\ref{tmatrix}) can be systematically made if necessary, e.g. by better dealing with
the poles $q_n$ in Eq.(\ref{gammak}) and the constants $\alpha_l$.

\section{Confinement induced effects}
\label{resonantscatt}
We now use the above formalism to analyze confinement induced phenomena that occur both in
the scattering by a central field and in collision processes. Although related to each
other, two important phenomena can be distinguished: resonant scattering
(sections~\ref{bffb} and~\ref{cirf0}) and weakly localized states induced by the
confinement (section~\ref{bstates}). In the present paper, our formalism is focused on the
single mode regime. It is 
shown also that previous results of atom-atom collisions under parabolic confinement are
recovered, regarding not only the scattering resonances
(CIR)~\cite{olshanii1998a,moore2004a,granger2004a} but also the confinement induced weakly 
bound states~\cite{bergeman2003}. CIR in atomic collisions involving excited transversal
states at energies $k^2>q_1^2\,$ are treated in~\cite{moore2004a} for interactions in the
$s$-wave approximation and in~\cite{granger2004a} for general $V(r)\,$. See also
independent related works for 2DEG mesoscopic
systems~\cite{chu1989a,bagwell1990a,gurvitz1993a,bardarson2004a}.

\subsection{Single mode regime}
\label{smode}
In this regime, the total scattering energy $E=\hbar^2k^2/2\mu$ allows for only
the transversal ground state $\varphi_0(\rho)$ to be effectively occupied
$k^2=q_0^2+k_0^2<q_1^2$, i.e., to be an open channel. All excited states can only be
virtually populated due to the scattering of the incoming state. Then $n_E=0$ and
$b_n=\delta_{n,0}$. 
In terms of the even $f_{0g}$ and odd $f_{0u}$ angular momenta decomposition of $f_0^\pm$
given in Eq.(\ref{f1d-spherical}), namely $f_0^\pm=f_{0g}\pm f_{0u}$, the current
conservation can be written as
\begin{equation}
\label{conservationfgu}
\left(\mathrm{Re}\{f_{0g}\} + |f_{0g}|^2\right) 
	+ \left(\mathrm{Re}\{f_{0u}\} + |f_{0u}|^2\right) = 0\, .
\end{equation}
A suitable parametrization of $f_{0g,u}$ is obtained by introducing 1D scattering
parameters $\delta_g$ and $\delta_u$, such that 
\begin{subequations}
\label{sectors-normalized}
\begin{eqnarray}
\label{sectors-normalized-g}
f_{0g} &\equiv& - \frac{1}{1 + i\cot{\delta_g}}\, , \\
\label{sectors-normalized-u}
f_{0u} &\equiv& - \frac{1}{1 + i\cot{\delta_u}}\, .
\end{eqnarray}
\end{subequations}
In this form, Eq.(\ref{conservationfgu}) is verified if $\delta_{g,u}$ are
arbitrary real numbers, since then $\mathrm{Re}\{f_{0g,u}\} + |f_{0g,u}|^2=0$ vanishes
separately. There is by now no {\em a priori} reason for the parameters $\delta_{g,u}$ to
be both real. However, in the case of general potentials $v(r)$ and $u(\rho)$ considered
in this work, it will be seen in the next sections that Eqs.(\ref{sectors-normalized})
with real $\delta_{g,u}$ provide a valuable means to establish the boson-fermion and
fermion-boson mappings as well as the conservation condition
Eq.(\ref{conservationfgu}).

\subsection{Boson-fermion and fermion-boson mappings}
\label{bffb} 
Consider now a collisional process between \emph{identical} particles. It is convenient to 
split $\Psi(\bm{r})$ into symmetric $\psi_g(z)\varphi_0(\rho)$ and antisymmetric
$\psi_u(z)\varphi_0(\rho)$ parts, with respect to $\bm{r}\rightarrow -\bm{r}$,
\begin{eqnarray*}
\Psi(\bm{r}) &=& \left[e^{ik_0z} + f_0^\pm\ e^{ik_0|z|}\right]\varphi_0(\rho)\, \\
		&=& \left[\psi_g(z) + \psi_u(z)\right]\,\varphi_0(\rho)\, ,
			\hspace{2em} |z|>R_{1D}'\, .
\end{eqnarray*}
$\psi_g(z)$ and $\psi_u(z)$ can be deduced by using the form Eq.(\ref{f1d-spherical})
for the amplitude $f_0^\pm$ in terms of its even $f_{0g}$ and odd $f_{0u}$ sectors. For
the symmetric part, we obtain, 
\begin{subequations}
\label{sectors}
\begin{eqnarray}
\label{sectors-g}
\hspace{-2em}
\psi_g(z) &=& (1+f_{0g})\cos{(k_0z)}+if_{0g}\sin{(k_0|z|)}	\nonumber\\
	  &=& e^{i\delta_g}\cos{(k_0|z| + \delta_g)}\, ,
			\hspace{2em} |z|>R_{1D}'\, ,	
\end{eqnarray}
where Eq.(\ref{sectors-normalized-g}) has been employed. The antisymmetric part is given
by 
\begin{eqnarray}
\label{sectors-u}
\hspace{-2em}
\psi_u(z) &=& i(1+f_{0u})\sin{(k_0z)}\pm f_{0u}\cos{(k_0z)}	\nonumber\\
	  &=& ie^{i\delta_u}\sin{(k_0z \pm \delta_u)}\, ,
			\hspace{2em} |z|>R_{1D}'\, ,
\end{eqnarray}
\end{subequations}
where the plus (minus) sign in $\psi_u(z)$ refers to $z>0$ ($z<0$) and
Eq.(\ref{sectors-normalized-u}) has been used.   

If the colliding particles are e.g. spin polarized $\Psi(\bm{r})$ must be (anti-)
symmetrized. For bosons, the correct effective 1D scattered wavefunction in the asymptotic
region $|z|>R_{1D}'$ is $\psi_g(z)$. The {\em resonance} condition can be identified as 
$f_{0g} = -1$
or $\delta_g=|\pi/2|$
and it follows from Eq.(\ref{sectors-g}) that the wavefunction becomes
$\psi_g(z)=-i\sin{(k_0|z|)}$, which is with respect to its modulus the wavefunction of a
pair of {\em non-interacting} fermions, as can be seen by setting $f_{0u}\equiv 0$ (or 
$\delta_u\equiv 0$) in Eq.(\ref{sectors-u}). This is the well-known fermionic mapping of 
strongly interacting, inpenetrable bosons in
1D~\cite{tonks1936,girardeau1960,lenard1966,paredes2004a}. The bosons would not be allowed
to be located at the same position $z=0$, supposing the asymptotic solution 
Eq.(\ref{sectors-g}) could be extended towards the origin $|z|\ll R_{1D}'\,$. 

For fermions, a reciprocal mapping exists at resonance 
\begin{equation}
\label{identical-u}
f_{0u} = -1\,, \hspace{3em} \mathrm{or} \hspace{3em} \delta_u=|\pi/2|\, .\\
\end{equation}
From Eq.(\ref{sectors-u}), the wavefunction becomes $\psi_u(z)=\mp\cos{(k_0z)}$, which is
the wavefunction of a pair of {\em non-interacting} bosons~\cite{granger2004a}. Clearly,
from the exact, fully antisymmetric wavefunction of a pair of spin polarized fermions, the
probability of finding both of them at the origin $\bm{r}=0$ is strictly zero. The
bosonization implied by this confinement induced resonance means that the asymptotic
behavior of the fermions far from the interaction region is that of free bosons. 

These mappings between identical colliding particles are valid also under a {\em
longitudinal confinement} along the $z$-axis. In order to neglect couplings to the center
of mass, this longitudinal confinement can be chosen to be parabolic. The characteristic
oscillator length $R_\parallel$ should then be flat enough, such that $R_\parallel\gg
R_{1D}'$. The asymptotics of the symmetric and antisymmetric eigenstates with (positive)
energies $E_{g,u}=\hbar^2k_{g,u}^2/2\mu$ are given by Eq.(\ref{sectors-g}) and
Eq.(\ref{sectors-u}), respectively. The corresponding eigenvalues 
$k^2_{g,u}=q_0^2+k_{0g,u}^2$ can be estimated by solving the equations 
$\psi_{g,u}(R_\parallel)\approx 0$ for $k_{0g,u}$, namely, 
		$\cos{(k_{0g}R_\parallel + \delta_g)} = 0$ and 
		$\sin{(k_{0u}R_\parallel + \delta_u)} = 0$.
If the parameter $\delta_g=\delta_g(k_{0g})$ is made equal to $\pm\pi/2$ by varying, e.g.,
$R_V$, $R_U$, or $R_\parallel$, one sees that $\psi_g(z)$ is once more mapped onto an 
antisymmetric eigenstate without interaction $v(r)\equiv 0$ (or $\delta_u\equiv 0$), as
already seen above for free longitudinal motion. Analogously, $\psi_u(z)$ is mapped onto a 
non-interacting symmetric eigenstate as $\delta_u=\delta_u(k_{0u})\rightarrow
\pm\pi/2\,$. This latter type of bosonic mapping of fermionic eigenstates under
longitudinal confinement was first verified numerically in~\cite{granger2004a} for 
the collision interaction $V(r)=d/\cosh{(r/b)}^2\,$. 

Note that, if one is not restricted to the case of collisions, this result can be formally
quite general regarding not only the interaction potential $v(r)\,$, but also some
sufficiently flat transversal and longitudinal confinements.

\subsection{The effective amplitude $f_0^\pm$}
\label{f0pm}
The considerations in the preceding section assumed the general form
Eqs.(\ref{sectors-normalized}) for $f_{0g,u}\,$. In this section, $f_0^\pm$ is explicitly 
calculated by solving the matrix equation Eq.(\ref{tmatrix}) for the elements $T_l\,$.  
From $k\sim 1/R_U$, it follows that $kR_V\sim R_V/R_U\ll 1$, which is the condition of
low scattering energy in 3D free space scattering and the phase-shifts 
\[
\tan{\delta_l}=\tan{\delta_l(k)}\sim k^{2l+1}\sim 1/R_U^{2l+1}
\] 
are generally small~\cite{mott1965} for large $R_U$ (or small $R_V$). One expects then
that orbital angular momenta higher than the leading contributions, e.g. the $s$- and
$p$-waves, should not significantly change the main features arising from the latter. 
Indeed, the calculation of $f_{0g}$ and $f_{0u}$ can be done separately, since the even
and odd angular momenta in Eq.(\ref{tmatrix}) are uncoupled from each other. In the single
mode regime, Eq.(\ref{gammak}) becomes  
$\gamma = 2/d_U^2\,k\,k_0$
and
$d_U\equiv R_U/N_0$,
where $k_0=(k^2-q_0^2)^{1/2}$ and $f_0^\pm$ in Eq.(\ref{f1d-spherical}) depends only on the
ratio $t_l\equiv T_l/k_0$. The matrix equation Eq.(\ref{tmatrix}) in this case can be
rewritten in terms of these $t_l\,$. Thus for all $l\,$, 
\begin{eqnarray*}
\lefteqn{\hspace{-5em}
\left\{2i - \left[k\cot{\delta_l} + (2l+1)P_{ll}\right]d_U^2k_0\right\}t_l  
	}								\nonumber\\
& = & \alpha_l\,d_U^2 + k_0\,d_U^2\sum_{s[l]\neq l} (2s+1)P_{ls}\,t_s\, , \hspace{-5em}
\end{eqnarray*}
where the diagonal term $s[l]=l$ is excluded from the series on the rhs. Assume now in
this equation for $t_l$ that this series converges. As $l$ increases,
$\alpha_l=\alpha_{l0}=P_l(k_0/k)/\pi^{1/2}d_U$ (see Eq.(\ref{spherical}) and
Eq.(\ref{alphaln})) decrases, while $P_{ls}$ (see Eq.(\ref{gammal})) should decrease
mildly due to the oscillation of $P_l(ix)\,$. Then the rhs of this equation for $t_l$
should decrease relatively smoothly with $l\,$. On the other hand, the phase $\delta_l$ 
decreases exponentially $\delta_l\sim 1/R_U^{2l+1}$ with $l$ at low energies, so that
$\cot{\delta_l}$ increases exponentially. Then $t_l$ on the lhs should be exponentially
small with increasing $l\,$. This is consistent with the assumption of convergence of the
series on the rhs [see also discussion on $\gamma_l(z)$ after Eq.(\ref{spherical})]. Apart
from this, one can use the smallness of the phase-shifts $|\delta_2|\ll |\delta_0|$ and 
$|\delta_3|\ll |\delta_1|$ in order to explicitly diagonalize the finite subspaces $l=0,2$
and $l=1,3$ separately. Then the couplings to $l=2$ and $l=3$ are seen to have a minor
effect to the leading terms $l=0$ and $l=1$, respectively. Hence we put
\begin{equation*}
t_l\approx \frac{\frac{1}{2}d_U^2\alpha_l}
	{i - \left[k\cot{\delta_l} + (2l+1)P_{ll}\right]\frac{1}{2}d_U^2k_0} 
							\,, \hspace{1em} l=0,1,\dots\, .
\end{equation*}
The even sector $f_{0g}$ is then given by the leading term, e.g, the $s$-wave, for which 
$k\cot{\delta_0}\approx -1/a\,$, where $a$ is the $s$-wave 3D free space scattering
length~\cite{mott1965}. From Eq.(\ref{f1d-spherical}) one obtains  
\begin{subequations}
\label{f0gu}
\begin{eqnarray}
\label{f0g}
f_{0g} &\equiv& \sum_{l=0,2,\dots}^\infty \frac{4\pi\,(2l+1)\,\alpha_{l0}}{2i}\ t_l
									\nonumber\\
&\approx& - \frac{1}{1+i\left[-\,\frac{d_U^2}{2a}\left(1-aP_{00}\right)\right]k_0 }\,.
\end{eqnarray}
The odd sector $f_{0u}$ is approximated by e.g. the $p$-wave. In this case, the relevant
quantity~\cite{suno2003a} is the scattering volume $V_p$ related to $\delta_1$ by  
$k\cot{\delta_1}\equiv -1/V_pk^2\,$. Then 
\begin{eqnarray}
\label{f0u}
f_{0u} &\equiv& \sum_{l=1,3,\dots}^\infty \frac{4\pi\,(2l+1)\,\alpha_{l0}}{2i}\ t_l
									\nonumber\\
&\approx& -\frac{3[P_1(k_0/k)]^2}
	{1+i\left[-\frac{d_U^2}{2}\left(\frac{1}{V_pk^2}-3P_{11}\right)\right]k_0 }\,. 
\end{eqnarray}
\end{subequations}

The range of validity of these results can be obtained by imposing e.g. the conservation
of probability. As discussed in connection to Eq.(\ref{conservationfgu}) and 
Eqs.(\ref{sectors-normalized}), the 1D scattering phase-shift $\delta_g$ is real for the
result Eq.(\ref{f0g}). Then a sufficient condition for Eq.(\ref{conservationfgu}) to hold
is 
\begin{subequations}
\label{k0}
\begin{equation}
\label{k0small}
|k_0|\ll k\sim 1/R_U 	
\end{equation}
since then $3[P_1(k_0/k)]^2\ll 1\,$ so that $f_{0u}\approx 0$ and the parameter
$\delta_u\approx 0$ can be considered also real. In this first region of low longitudinal
momenta, the angular momenta couplings in the matrix equation for $t_l$ should become
negligible due to the factor $k_0$ that multiplies the series $d_U^2\sum_{s[l]\neq
l}(2s+1)P_{ls}t_s$ containing other angular momenta. The present uncoupled solution to
$t_l$ should then be a good approximation. The second region is at relatively large
momenta, when $3[P_1(k_0/k)]^2\sim 1$, i.e., when 
\begin{equation}
\label{k0large}
|k_0|\sim q_0/\sqrt{2}\sim 1/R_U	
\end{equation} 
\end{subequations}
since then $\delta_u$ can also be taken as real. The decoupling between odd angular
momenta rests then on the smallness of the phase-shifts $\delta_l\,$. These are,
nevertheless, the most interesting regions for the momentum $k_0\,$ that are relevant to
resonant scattering in symmetric and antisymmetric states, respectively. For more detailed 
applications these regions can be enlarged by improving on Eq.(\ref{tmatrix}).

\subsection{Confinement induced resonances - CIR}
\label{cirf0}
As presented in sections~\ref{smode} and~\ref{bffb} in general terms, the CIR is one
remarkable effect arising due to the presence of a strong confinement. In the following,
it is explicitly expressed in terms of the parameters determining a given scattering 
potential and a given confinement.

\subsubsection{Even angular momenta}
\label{bosons}
The symmetric wavefunction in the single mode regime is $\psi_g(z)\varphi_0(\rho)$ and
only the symmetric sector $f_{0g}$ contributes. It can be used to describe
e.g. spin-polarized bosons or fermions in an antisymmetric spin state. In the expression
Eq.(\ref{f0g}) for this $f_{0g}$, $P_{00}$ is given in Eq.(\ref{gammal}), thus  
\begin{eqnarray}
\label{p00}
P_{00} &=& k\int_0^{p_c/k}dx\, P_0(ix)P_0(ix) \nonumber\\
	&=& \sqrt{q_1^2 - k^2}\, ,
\end{eqnarray}
having used the definition of $p_c$ in Eq.(\ref{deltac}) and the single mode condition
$n_E=0\,$. 

At \emph{small} scattering lengths $|a|\ll R_U\,$, one can neglect this factor $P_{00}$ in 
Eq.(\ref{f0g}), since then $|a|P_{00}\ll 1\,$, and 
\begin{eqnarray*}
f_{0g} 
&\approx& - \frac{1}{1+i\left[-\,\frac{d_U^2}{2a}\right]k_0}\,, 
						\hspace{2em} |a|\ll R_U\, .
\end{eqnarray*}
This is also true at ``high'' energies $|k_0|$ when $k\rightarrow q_1$ such that 
$P_{00}\approx 0$. One must, however, be careful about the validity of the present
approximations in this region of large momenta $|k_0|$, as mentioned above in the context
of Eq.(\ref{k0large}). The only possibility for resonance $f_{0g}=-1$ is then at zero
energy $k_0=0\,$.  

At \emph{large} scattering lengths $R_V\ll |a|\sim R_U\,$, this factor $P_{00}$ plays an 
important role in the scattering process at low momenta $|k_0|\ll 1/R_U$ for which  
$P_{00}\approx\sqrt{q_1^2-q_0^2}$ is maximum. Indeed, one obtains then, for 
$|k_0|\ll 1/R_U\,$,  
\begin{subequations}
\label{f0gc'}
\begin{equation}
\label{f0glargea}
f_{0g} \approx - \frac{1}{
		1+i\left[-\,\frac{d_U^2}{2a}\left(1-\frac{C'a}{d_U}\right)\right]k_0 
		   }\,,			
\end{equation}
where the constant $C'$ depends \emph{only} on the first closed and the last open
channels, namely  
\begin{eqnarray}
\label{c'}
C' &\equiv& d_U\sqrt{q_1^2 - q_0^2}			\nonumber\\
   & = &    \frac{1}{(\hbar^2/2\mu d_U^2)^{1/2}}\,\sqrt{\epsilon_1 - \epsilon_0}\, .
\end{eqnarray}

Recalling that $f_0^\pm\approx f_{0g}$, since $f_{0u}\approx 0$ for such small momenta
$|k_0|$ (see Eqs.(\ref{f0gu})), this result implies that the scattering can be described
by means of an effective 1D interaction potential $V_{1D}(z)$ of zero range 
\begin{equation*}
V_{1D}(z) = g_{1D}\delta(z)\,, 
\hspace{1.5em} 	g_{1D} = \frac{\hbar^2}{\mu}\frac{2a}{d_U^2}\frac{1}{1-C'a/d_U}\,.
\end{equation*}
\end{subequations}
In other words, Eq.(\ref{f0glargea}) is the solution to $f_0^\pm$ for the hypothetical
pure 1D scattering problem under this potential $V_{1D}(z)$ and is valid also for
distinguishable particles.  

As discussed in section~\ref{cirpseudo} for atom-atom collisions, this same 1D potential
$V_{1D}(z)$ at low momenta $|k_0|$ was first shown in~\cite{olshanii1998a} to be a direct 
result of parabolic confinement and of the 3D zero range interaction potential
$V_\delta(r)$ of Eq.(\ref{pseudo}). The occurence of this collisional resonance at
$a_\perp=Ca$ for general interactions $V(r)$ was then shown
in~\cite{granger2004a}. 

For other types of scattering processes, e.g. in mesoscopic 2DEG 
systems~\cite{chu1989a,bagwell1990a,gurvitz1993a,bardarson2004a}, Eq.(\ref{f0glargea})
shows therefore that at low longitudinal momenta $k_0$ the existence of a resonance of the
type 
$f_0^+\approx f_{0g}= -1$
is characterized by an infinite effective 1D coupling strength 
\begin{equation}
\label{cir}
|g_{1D}|\rightarrow\infty\, , \hspace{3em} d_U\approx C'a\, ,
\end{equation}
and is indeed not restricted to atom-atom collisions under parabolic
confinement or to two-dimensional quantum point-contacts with a single
impurity~\cite{gurvitz1993a}. It is a general effect of single mode scattering at low
velocities under quasi-1D confinement.

\subsubsection{Odd angular momenta}
\label{oddcir}
For odd angular momenta, the procedure is similar to the previous analysis. The
antisymmetric part $\psi_u(u)$ shows a scattering resonance when the associated amplitude 
$f_{0u}=-1\,$. At this resonance, the system can be mapped to a non-interacting symmetric
wavefunction as discussed in connection to Eq.(\ref{identical-u}). The term $P_{11}$ is
given in Eq.(\ref{gammal}), from which the actual condition of CIR in the amplitude
$f_{0u}$ in Eq.(\ref{f0u}) can be worked out for given scattering and confining
potentials. Note that due to the factor $3[P_1(k_0/k)]^2$ in the numerator of
Eq.(\ref{f0u}), the odd component $f_{0u}$ is non-negligible only at relatively large
longitudinal momenta $|k_0|\sim 1/R_U$, in contrast to the even amplitude $f_{0g}$
discussed above. 

For atom-atom collisions under parabolic confinement, the existence of CIR beyond the 
$s$-wave was first predicted by neglecting couplings to orbital angular momenta and
isolating a single partial wave $l>0$ and its phase-shift $\delta_l$ as the sole dominant
contribution~\cite{granger2004a}. A direct consequence of contributions from non-zero
partial waves $l>0$ is, among others, the possibility to define effective $p$-wave
\emph{zero-range} scattering potentials analogous to $V_\delta(r)$ and $V_{1D}(z)$, but
which act only on antisymmetric wave functions of a colliding
pair~\cite{kanjilal2004a,girardeau2004a}.

\subsection{Weakly bound states}
\label{bstates}
Besides CIR, a second effect of strong cylindrical confinement is the prediction of weakly
bound and localized states. For atom-atom collisions this implies the formation of 
quasi-molecules. As first shown in~\cite{bergeman2003} for zero-range atom-atom interactions
under parabolic confinement, these bound states exist even when no bound state occurs in 
free space. They have recently been observed in optical traps loaded with $^{40}$K 
atoms~\cite{moritz2005a}. Similar bound states localized around impurities are found in
independent studies of mesoscopic systems~\cite{chu1989a,bagwell1990a,gurvitz1993a,
bardarson2004a}. In this section, a localized state for general cylindrical confinement is 
calculated from the even, more precisely zero, angular momentum sector.

\subsubsection{$s$-wave binding energy and wavefunction}
For the \emph{free space} $l=0$ angular momentum, the bound state of the (attractive)
scattering potential $V(r)$ has an energy $E_{Bf}$ given by  
\[
E_{Bf}\equiv -\frac{\hbar^2\kappa_{Bf}^2}{2\mu} \approx -\frac{\hbar^2}{2\mu\,a^2}\, ,
				\hspace{2em} a\gg R_V
\]
whose relationship to the scattering length $a$ via $\kappa_{Bf}\approx  1/a$ holds only
when $a\gg R_V>0$~\cite{mott1965}. At negative scattering lengths $a<0$, one can speak of
a ``virtual'' state close to be incorporated to the spectrum of $V(r)$. Under lateral
confinement and along the radial direction of the cylindrical trap, the tail of its
wavefunction $e^{-\kappa_{Bf}r}$ is changed to vanish at a finite distance from the
origin, i.e. at the edge $r=\rho=R_U$ of the guide when $|a|\sim R_U$ (see
Fig.~\ref{guideeigenstates}). This lateral squeeze lifts $E_{Bf}<0$ by an amount
$\epsilon_0$. It can be sufficient for this state to pass the threshold without
confinement $E=0$ as $R_U$ decreases further.  

This confined localized state with energy $E_B$ satisfies Eq.(\ref{hrelativ2}) with  
$k^2$ replaced by $2\mu E_B/\hbar^2$. This replacement is equivalent to redefining $k_0$
to be the imaginary number $k_{0B}\equiv \pm\, i(q_0^2-2\mu E_B/\hbar^2)^{1/2}$.		
Since the wave $e^{ik_{0B}z}$ should be absent from the solution $\Psi$ and 
$e^{ik_{0B}|z|}$ should show a bound state like exponential decay, $1/f_0^\pm(k_{0B})$ must
vanish at a \emph{positive} imaginary value of $k_{0B}\,$, such that the interaction
part $e^{ik_{0B}|z|}$ outweights $e^{ik_{0B}z}$. Let then $k_{0B} \equiv ix_B/a$, so that
$x_B>0$ if $a>0$ and $x_B<0$ if $a<0\,$. In terms of this $x_B\,$, $E_B$ can be rewritten as 
\begin{equation}
\label{bse}
E_B \equiv \frac{\hbar^2}{2\mu}\left( q_0^2 - \frac{x_B^2}{a^2}\right), 
			\hspace{2em}  a\,x_B>0\, .
\end{equation}
From Eqs.(\ref{f0gu}), a \emph{root} of $1/f_0^\pm$ arises either from the even sector
$1/f_{0g}$ or from the odd sector $1/f_{0u}\,$. In order to recover the $l=0$ free space
state $E_{Bf}$, this root should come from the even sector Eq.(\ref{f0g}). Using then the
substitution $k^2\rightarrow 2\mu E_B/\hbar^2=q_0^2 - x_B^2/a^2$ in the expression
Eq.(\ref{p00}) for $P_{00}$, the equation for $x_B$ becomes
\begin{subequations}
\label{xbquartic}
\begin{equation}
\label{xb}
2s^2  + \left[ 1 - \frac{a}{|a|} \sqrt{(s\,C')^2 + x_B^2}\right]x_B = 0\, ,
		\hspace{1em} s\equiv\frac{a}{d_U},
\end{equation}
where $|a|$ arises as $1/a^2$ is taken out of the square root in $P_{00}\,$. Another 
enlarged form is the quartic equation 
\begin{equation}
\label{quartic}
x_B^4 + \left(C'^2s^2 - 1\right)x_B^2 - 4s^2x_B -4s^4 = 0\, .
\end{equation}
\end{subequations}
For each $|s|$, one should pick up the correct positive (negative) root $x_B$ if $s>0$
($s<0$), as indicated in Eq.(\ref{bse}). 

\begin{figure}
\includegraphics[scale=0.22]{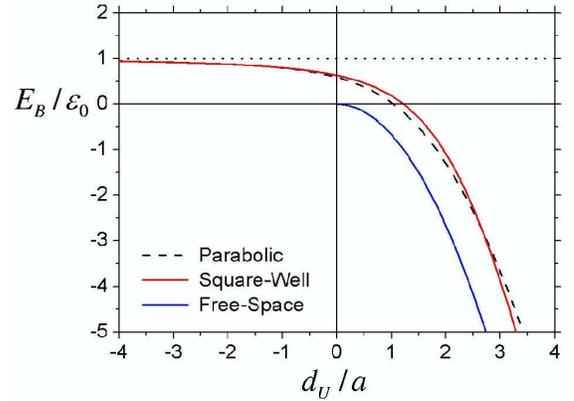}
\caption{\label{boundstates} (Color online) $s$-wave binding energy $E_B$ for parabolic (dashed 
curve) and square well (upper solid curve) confinement, scaled by the respective ground state
energies $\epsilon_0$. The free space energy (lower solid curve) $E_{Bf}$ is scaled with
the square well value of $\epsilon_0$ and is defined only for positive $a$. The dotted
line marks the continuum threshold under confinement.}
\end{figure}

The weak confinement limit is $|s|\rightarrow 0\,$. A finite root $x_B\approx +1$ exists 
only for $a>0\,$ as can be seen from Eq.(\ref{xb}). This means that 
$E_B\approx E_{Bf}$, 
when $R_U\rightarrow \infty$ or when $a\rightarrow 0^+$ as expected. For $a<0\,$, a trial
Taylor expansion $x_B\approx x_1s + x_2s^2$ in Eq.(\ref{quartic}) yields $x_1=0$ and
$x_2=-2$. A localized state with energy 
\begin{equation*}
E_B\approx \epsilon_0 - 4\left(\frac{a}{d_U}\right)^4|E_{Bf}|, \hspace{2em} |a|\ll d_U\, ,
\end{equation*}
is thus expected also for negative scattering lengths. In this case, the confinement has
been able to turn the``virtual'' state $E_{Bf}$ of $V(r)$ into a real bound state
localized around the scattering center. 

The strong confinement limit is $|s|\gg 1\,$ (assuming the basic requirement $R_V\ll R_U$
is not violated). Setting then $x_B\approx x_\infty/ s^{-1}$ in Eq.(\ref{quartic}) and
reexpressing it in terms of $s^{-1}$ gives for positive and negative $a\,$, respectively,  
$x_\infty = \pm \{[(16 + C'^4)^{1/2} - C'^2]/2\}^{1/2}$. 
For both positive and negative roots $x_B\,$, the same fraction of the transversal 
ground state energy $\epsilon_0$ is reached as $|s|\rightarrow\infty$
\begin{equation*}
E_B\approx \left[1 - \left(\frac{x_\infty}{q_0d_U}\right)^2\right]\epsilon_0\, ,
				\hspace{2em} |a|\gg d_U\, .
\end{equation*}
For parabolic guide $d_U=a_\perp$ [see Eq.(\ref{gammak}) with the exact value
$|\varphi_n(0)|^2=1/\pi a_\perp^2$]. Then $C'=2$ and $q_0a_\perp=\sqrt{2}$, both obtained
by directly using the exact energies $\epsilon_0\equiv
\hbar^2q_0^2/2\mu=\hbar\omega_\perp\,$ and $\epsilon_1\equiv
\hbar^2q_1^2/2\mu=3\hbar\omega_\perp\,$ in terms of $\omega_\perp$. 
Hence \mbox{$E_B = (2-\sqrt{2})\epsilon_0 = 0.586\epsilon_0\,$} in good quantitative
agreement with~\cite{bergeman2003,moritz2005a}. For square well confinement, it follows
from Eq.(\ref{besselroot}) and Eq.(\ref{c'}) that $C'=(20/3)^{1/2}=2.582$ and
$q_0d_U=(3/2)^{1/2}=1.225$, thus $E_B = 0.631\epsilon_0\,$. These energies are plotted in
Fig.~\ref{boundstates} as function of the confinement length scale $d_U$.

The spatial shape $\Psi_B(\bm{r})$ of this confinement induced weakly bound state in the
region \emph{far} from the scattering center is 
\begin{subequations}
\label{PsiB}
\begin{equation}
\label{PsiB-asy}
\Psi_B(\bm{r})\propto e^{-x_B|z|/a}\,\varphi_0(\rho), \hspace{3em} |z|\gg R'_{1D}\,,
\end{equation}
setting aside overall constants and where $x_B$ is the proper root of
Eq.(\ref{xbquartic}). This asymptotic form is common to both positive and negative
scattering lengths $a$, with $a\,x_B>0$. \emph{Closer} to the origin, the wavefunction can
be calculated by analytic continuation of the intermediate asymptotics in
Eq.(\ref{spherical}) and Eq.(\ref{gammal}). There the function $T_0=T_0(k_0)$ diverges at
$k_0=i\,x_B/a$, which is the condition for the pole of $f_0^\pm$. After this analytic
continuation, the step towards Eq.(\ref{spherical-delta}) can no longer be taken (it was
taken in order to calculate $T_0=T_0(k_0)$ for real $k_0$). In this series
Eq(\ref{spherical}), this divergence then singles out terms proportional to the constant
$T_0(i\,x_B/a)\rightarrow\infty\,$. As a result, apart from overall constants, the
wavefunction $\Psi_B(\bm{r})$ has the following form (for $R_V\ll r\ll R_U$) 
\begin{eqnarray}
\label{PsiB-intasy}
\lefteqn{\Psi_B(\bm{r})} 							\\
&\propto& \left[\sqrt{q_1^2-\kappa_B^2}
	- \frac{2/d_U^2}{\sqrt{q_0^2-\kappa_B^2}}\right] 
			j_0(\kappa_Br) + \kappa_B\,n_0(\kappa_Br), 	\nonumber
\end{eqnarray}
\end{subequations}
where $\kappa_B^2\equiv 2\,\mu E_B/\hbar^2$ and $E_B$ is given in Eq.(\ref{bse}). For
$E_B<0$, it can be seen that both imaginary values of $\kappa_B$ yield the same result. In 
particular, the free space bound state is recovered in the limit $a/d_U\rightarrow
0^+$. Using $E_B\approx E_{Bf}\approx -\hbar^2/2\mu a^2$ in Eq.(\ref{PsiB-intasy}), one
has  
\begin{equation*}
\Psi_B(\bm{r})\propto - \frac{e^{-r/a}}{r}\,, \hspace{3em} a/d_U\rightarrow 0^+. 
\end{equation*}
For negative scattering length $a/d_U\rightarrow 0^-$, the free space ``virtual'' bound
state turns into a real one whose energy was calculated in the previous section, $E_B
\approx \epsilon_0 - 4(a/d_U)^4|E_{Bf}|$. The wavefunction Eq.(\ref{PsiB-intasy}) becomes
instead 
\begin{equation*}
\Psi_B(\bm{r})\propto - \frac{1}{|a|}\,j_0(q_0r) + q_0n_0(q_0r)\,, 
						\hspace{2em} a/d_U\rightarrow 0^-. 
\end{equation*}
On the opposite limit of strong confinement $|a|/d_U\rightarrow\infty$, $E_B$ tends to a
positive fraction $\lambda$ of $\epsilon_0$, and the wavefunction becomes a superposition
of both terms $j_0(\sqrt{\lambda}\,q_0r)$ and $n_0(\sqrt{\lambda}\,q_0r)$.

\subsubsection{Higher angular momenta}
Localized states from higher angular momenta contributions and their \emph{couplings}
among each other can in principle be systematically obtained by solving Eq.(\ref{tmatrix})
and analyzing the poles of the full amplitude $f_0^\pm=f_{0g}\pm f_{0u}\,$. For instance,
a $d$-wave pole should come from retaining the $l=2$ contribution. This and other poles
from the odd sector $f_{0u}$ are treated in detail elsewhere.

\section{Discussion}
\label{harmonic}
Here the role of the continuum approximation in section~\ref{gfunctions} is discussed by
comparing with previous models for scattering under confinement. In the interior of the
guide $r,r'\ll R_U$ where the confining potential is negligible, the Green's function
$G_c(\bm{r},\bm{r}')$ under confinement approaches the 3D free space Green's function
$G(\bm{r},\bm{r}')$, as shown in Eq.(\ref{gcgu2}). Therefore, the singularity of
$G_c(\bm{r},\bm{r}')$ as $r,r'\rightarrow 0$ is not only coincident with (see
section~\ref{cirpseudo}), but it is essentially \emph{the} well-known singularity
$1/|\bm{r}-\bm{r}'|$ of the 3D scattering scenario without confinement. In addition, one
expects that the details of the confinement, whether parabolic or not, should not be
important for the physical understanding. In fact, the spectrum of \emph{excited}
transversal states sums up to yield $G(\bm{r},\bm{r}')$ (integrated over the axial angle
$\phi'$), as the continuum limit shows. The only condition is a short-ranged scattering
potential $R_V\ll R_U$ or a correspondingly ``flat'' confinement. 

The remarkable effects induced by the confinement are in turn directly related to the 
\emph{low lying} transversal states as also demonstrated in~\cite{gurvitz1993a} for a
two-dimensional problem. The discreteness of these states is then captured in the factor
$\gamma$ and in the correction term $\Delta_c(\bm{r},\bm{r}')$, which introduce
renormalizations and couplings of orbital angular momenta not accounted for by
$G(\bm{r},\bm{r}')$ alone.  

For a more quantitative analysis, consider the specific case of atom-atom collisions in
the low energy $s$-wave approximation and under parabolic
confinement~\cite{olshanii1998a,moore2004a,bergeman2003}. As discussed in 
section~\ref{cirpseudo}, the \emph{discrete} summation over the transversal states can be
dealt with in order to extract the singularity of the scattering solution and to determine
the value of the constant $C=1.4603\dots$ in the effective scattering amplitude
$f_0^\pm\,$. On the other hand, the corresponding value of $C$ in the \emph{continuum}
limit for this case is 
$C\equiv\mathrm{lim}_{s\rightarrow\infty}(\int_0^sds'/\surd{s'}-\sum_{s'=1}^s 1/\surd{s'})
\approx\int_0^1 ds'/\surd{s'}=2$ (see Eq.(9) of~\cite{olshanii1998a}), in agreement with
$C'=2$ calculated in section~\ref{bstates}. This value $C'=2$ arises by using the exact
harmonic oscillator energies $\epsilon_0=\hbar\omega_\perp$ and
$\epsilon_1=3\hbar\omega_\perp\,$. It can be improved on by e.g. numerically computing
$\Delta_c$. As already pointed out in section~\ref{gfunctions}, the exact excited spectrum
$\epsilon_n\sim 2n$ is substantially different from the ``flat'' potential approximation
given in Eq.(\ref{besselroot}). However, the discrete sum over this exact parabolic
spectrum can be shown~\cite{olshanii1998a,moore2004a} to have an equal singular behavior
$1/|z|$ as in free space (see section~\ref{cirpseudo}). Hence also within the discrete
approach, this demonstrates the major role played by the lowest transversal states,
whereas the sum over the excited states turns out to be qualitatively
confinement-independent.

\section{Conclusions and Perspectives}
\label{conclusions}
An analytical treatment of quasi-1D quantum scattering by spherically symmetric and 
short but finite range potentials in general cylindrical confinement is developed. The
full scattering wavefunction is calculated non-perturbatively without partially resuming 
perturbative series. All phase-shifts of the scattering potential can be readily
incorporated. This formalism provides a unified physical picture of the process of
confined quasi-1D scattering at low energies and can be applied to impurity scattering in
mesoscopic 2DEG systems and two-body collisions. Following the reasoning related to
Eq.(\ref{gcgu2}) one expects to be able to treat non-cylindrical geometries and quasi-2D
scattering in an analogous way. By computing then a few coefficients and functions such as
$\alpha_{ln}$ and $\Delta_c$ (or $\Delta_u$) and only the lowest eigenstates of the
confining potential, reasonable numerical results are also possible. Scattering resonances 
such as total reflexion, as well as weakly localized states can be determined. For the 
particular case of parabolic confinement, the formalism presented here can also be used to
obtain e.g. confinement induced two-body weakly bound states. Of particular interest is
the study of the poles of the scattering amplitude which correlate to orbital angular
momenta beyond \mbox{$s$-waves} and how non-parabolic geometries can affect these
unconventional pairings due to couplings to the center of mass coordinates.

\begin{acknowledgments}
J.I.K. appreciates financial support from the Conselho Nacional de Desenvolvimento Cient\'ifico 
e Tecnol\'ogico (CNPq) and the Alexander von Humboldt Foundation (AvH). J.S. thanks the DFG 
Schwerpunktprogramm: ``Wechselwirkung in Ultrakalten Atom- und Molek\"ulgasen'' for financial 
support.
\end{acknowledgments}


\begin{thebibliography}{99}
\bibitem{itrs} International Technology Roadmap for Semiconductors (ITRS), Annual Reports
(see http://public.itrs.net).
\bibitem{itrs-erd} See e.g. the ITRS chapter on Emerging Research Devices. 
\bibitem{weinstein1995a} J.D. Weinstein and K.G. Libbrecht, Phys. Rev. A {\bf 52}, 4004
(1995). 
\bibitem{folman2002a} R. Folman, P. Kr\"{u}ger, J. Schmiedmayer, \mbox{J. Denschlag} and
C. Henkel, Adv. At. Mol. Opt. Phys. {\bf 48}, 263 (2002). 
\bibitem{reichel2002a} J. Reichel, Appl. Phys. B {\bf 75}, 469 (2002). 
\bibitem{hinds1999a} E.A. Hinds and I.G. Hughes, J. Phys. D: Appl. Phys. {\bf 32}, R119
(1999). 
\bibitem{fortagh2003a} J. Fort\'{a}gh, H. Ott, S. Kraft, A. G\"{u}nther and
\mbox{C. Zimmermann}, Appl. Phys. B {\bf 76}, 157 (2003). 
\bibitem{wang2005a} Y.-J. Wang, D.Z. Anderson, V.M. Bright, \mbox{E.A. Cornell}, Q. Diot,
T. Kishimoto, M. Prentiss, R.A. Saravanan, \mbox{S.R. Segal} and S. Wu,
Phys. Rev. Lett. {\bf 94}, 090405 (2005). 
\bibitem{grimm2000a} R. Grimm, M. Weidem\"uller and Y.B. Ovchinnikov,
Adv. At. Mol. Opt. Phys. {\bf 42}, 95 (2000). 
\bibitem{oberthaler2003a} M.K. Oberthaler and T. Pfau, J. Phys.: Condens. Matt. {\bf 15},
R233 (2003). 
\bibitem{datta1997a} S. Datta, \emph{Electronic Transport in Mesoscopic Systems}
(Cambridge Uni. Press, Cambridge, 1997). 
\bibitem{ferry1997a} D.K. Ferry and S.M. Goodnick, \emph{Transport in Nanostructures}
(Cambridge Uni. Press, Cambridge, 1997).
\bibitem{chu1989a} C.S. Chu and R.S. Sorbello, Phys. Rev. B {\bf 40}, 5941 (1989).
\bibitem{bagwell1990a} P.F. Bagwell, Phys. Rev. B {\bf 41}, 10354 (1990).
\bibitem{gurvitz1993a} S.A. Gurvitz and Y.B. Levinson, Phys. Rev. B {\bf 47}, 10578
(1993).
\bibitem{bardarson2004a} J.H. Bardarson, I. Magnusdottir, \mbox{G. Gudmundsdottir},
C.S. Tang, A. Manolescu and V. Gudmundsson, Phys. Rev. B {\bf 70}, 245308 (2004). 
\bibitem{indlekofer2005a} K.M. Indlekofer, J. Knoch, and J. Appenzeller, arXiv:
cond-mat/0504746 at http://www.arxiv.org
\bibitem{tomonaga1950a} S.-I. Tomonaga, Prog. Theor. Phys. {\bf 5}, 544 (1950).
\bibitem{luttinger1963a} J.M. Luttinger, J. Math. Phys. {\bf 4}, 1154 (1963).
\bibitem{voit1994a} J. Voit, Rep. Prog. Phys. {\bf 57}, 977 (1994).
\bibitem{petrov2001a} D.S. Petrov, G.V. Shlyapnikov, and J.T.M. Walraven,
Phys. Rev. Lett. {\bf 87}, 050404 (2001).
\bibitem{goerlitz2001a} A. G\"orlitz, J.M. Vogels, A.E. Leanhardt, \mbox{C. Raman},
\mbox{T.L. Gustavson}, J.R. Abo-Shaeer, A.P. Chikkatur, \mbox{S. Gupta}, S. Inouye,
T. Rosenband and W. Ketterle, Phys. Rev. Lett. {\bf 87}, 130402 (2001).
\bibitem{dettmer2001a} S. Dettmer, D. Hellweg, P. Ryytty, J.J. Arlt, \mbox{W. Ertmer},
K. Sengstock, D.S Petrov, G.V. Shlyapnikov, \mbox{H. Kreutzmann}, L. Santos and
M. Lewenstein, Phys. Rev. Lett. {\bf 87}, 160406 (2001).  
\bibitem{tonks1936} L. Tonks, Phys. Rev. {\bf 50}, 955 (1936). 
\bibitem{girardeau1960} M. Girardeau, J. Math. Phys. (NY) {\bf 1}, 516 (1960). 
\bibitem{lenard1966} A. Lenard, J. Math. Phys. (NY) {\bf 7}, 1268 (1966).
\bibitem{tolra2004a} B. Laburthe Tolra, K.M. O'Hara, J.H. Huckans, \mbox{W.D. Phillips},
S.L. Rolston, and J.V. Porto, Phys. Rev. Lett. {\bf 92}, 190401 (2004).
\bibitem{paredes2004a} B. Paredes, A. Widera, V. Murg, O. Mandel, \mbox{S. F\"olling},
I. Cirac, G.V. Shlyapnikov, T.W. H\"ansch and \mbox{I. Bloch}, Nature {\bf 429}, 277
(2004).  
\bibitem{kinoshita2004a} T. Kinoshita, T. Wenger and D.S. Weiss, Science {\bf 305}, 1125
(2004). 
\bibitem{moritz2003a} H. Moritz, T. St{\"o}ferle, M. K{\"o}hl and T. Esslinger,
Phys. Rev. Lett. {\bf 91}, 250402 (2003).
\bibitem{moritz2005a} H. Moritz, T. St{\"o}ferle, K. G{\"u}nter, M.K{\"o}hl and
\mbox{T. Esslinger}, Phys. Rev. Lett. {\bf 94}, 210401 (2005). 
\bibitem{bostroem1981a} A. Bostr\"om and P. Olsson, J. Appl. Phys. {\bf 52}, 1187 (1981). 
\bibitem{olsson1994a} S. Olsson, Q. Jl. Mech. Appl. Math. {\bf 47}, 583 (1994). 
\bibitem{olshanii1998a} M. Olshanii, Phys. Rev. Lett. {\bf 81}, 938 (1998).
\bibitem{moore2004a} M.G. Moore, T. Bergeman, and M. Olshanii, J. Phys. IV France {\bf
116}, 69 (2004). 
\bibitem{granger2004a} B.E. Granger and D. Blume, Phys. Rev. Lett. {\bf 92}, 133202
(2004). 
\bibitem{bergeman2003} T. Bergeman, M.G. Moore, and M. Olshanii, Phys. Rev. Lett. {\bf
91}, 163201 (2003).    
\bibitem{petrov2000b} D.S. Petrov, M. Holzmann and G.V. Shlyapnikov, Phys. Rev. Lett. {\bf
84}, 2551 (2000). 
\bibitem{petrov2001b} D.S. Petrov and G.V. Shlyapnikov, Phys. Rev. A {\bf 64}, 012706
(2001). 
\bibitem{petrov2004a} D.S. Petrov, D.M. Gangardt and G.V. Shlyapnikov, J. Phys. IV France
{\bf 116}, 5 (2004). 
\bibitem{peano2005a} V. Peano, M. Thorwart, C. Mora and R. Egger, arXiv: cond-mat/{\bf
0506272} at http://www.arxiv.org 
\bibitem{effmass} A wave packet single-band effective mass approximation should be assumed
(see e.g~\cite{datta1997a,ferry1997a}).  
\bibitem{huang1987} K. Huang, \emph{Statistical Mechanics} (Wiley, New York, 1987). 
\bibitem{fetter1971a} A.L. Fetter and J.D. Walecka, \emph{Quantum Theory of Many-Particle
Systems} (McGraw-Hill, New York, 1971). 
\bibitem{morse1953} P. M. Morse and H. Feshbach, {\em Methods of Theoretical Physics}
(McGraw-Hill, 1953).   
\bibitem{mott1965} N. F. Mott and H. S. W. Massey, {\em The Theory of Atomic Collisions}
(Oxford Univ. Press, 1965). 
\bibitem{suno2003a} H. Suno, B.D. Esry and C.H. Greene, Phys. Rev. Lett. {\bf 90},
053202 (2003). 
\bibitem{kanjilal2004a} K. Kanjilal and D. Blume, Phys. Rev. A {\bf 70}, 042709 (2004). 
\bibitem{girardeau2004a} M.D. Girardeau and M. Olshanii, Phys. Rev. A {\bf 70}, 023608
(2004). 
\end{thebibliography}

\end{document}